\documentclass{article}
\usepackage{amsfonts, amsmath, amssymb, hyperref, graphicx, overpic}
\begin{document}
	\title{Product manifolds as realisations of general linear symmetries}
	\author{Tom Lawrence \\
	\it{\small{Ronin Institute for Independent Scholarship, 127 Haddon Place
		Montclair,}} \\
	\it{\small{New Jersey 07043-2314, United States of America}} \\
	\it{\small{tom.lawrence@ronininstitute.org}}}
\maketitle

\begin{abstract}
This paper considers the relationship between geometry, symmetry and fundamental interactions - gravity and those mediated by gauge fields. We explore product spacetimes which a) have the necessary symmetries for gauge interactions and four-dimensional gravity and b) reduce to an $N$-dimensional isotropic universe in their flat space limit. The key technique is looking at orbits of the operator form of symmetric rank-two tensors under changes of coordinate system. Orbits containing diagonal matrices are seen to correspond to product manifolds. The $GL(N,\mathbb{R})$ symmetry of the decompactified universe acts non-linearly on such a product spacetime.

We explore the resulting Kaluza-Klein theories, in which the internal symmetries act \emph{indirectly} on space of the extra dimensions, and give two examples: a six-dimensional model in which the gauge symmetry is $U(1)$ and a seven-dimensional model in which it is $SU(2)$.

We identify constraints that can be placed on any rank-two symmetric tensor to obtain such spacetimes: relationships between polynomial invariants. The multiplicities of its eigenvalues determine the dimensionalities of the factor spaces and hence the gauge symmetries.

If the tensor in question is the Ricci tensor, other than two-dimensional factor spaces all the factor spaces are Einstein manifolds. This situation represents the classical vacuum of the Kaluza-Klein theory.\footnote{\emph{This paper is a preprint of an article submitted for consideration in the \href{https://www.worldscientific.com/worldscinet/ijgmmp}{International Journal of Geometric Methods in Modern Physics} © (2022, World Scientific Publishing Company), modified in the following ways:
	\begin{itemize}
		\item This version does not use the journal's class file (formatting)
		\item It uses English rather than American English spellings and one typo/spelling error is corrected
		\item There is a minor clarification in appendix section A.2
		\item The Figures are of higher quality
		\item The text referring to one paper is improved
		\item For two papers in the Bibliography, hyperlinks to the articles on JSTOR are provided
	\end{itemize}
}}
\end{abstract}

\section{Introduction \label{Intro}}

\subsection{Motivation - a brief (and selective) history of unification}

Just under a century after Newton published his inverse square law of gravitation in 1687, Coulomb published his inverse square law of electrostatics. Over the following period of nearly 80 years, physicists examined electrical and magnetic phenomena and the interplay between them, culminating in Maxwell's theory of electromagnetism, showing how they were two parts of the same force. 

Hence at the start of the 20th century, the world of physics knew of two fundamental interactions, gravity and electromagnetism. They were described in similar ways: as deterministic field theories in three-dimensional space, with an absolute time. Over the next seventy years or so, all of this changed. Special relativity ended the idea of absolute time and theories became cast in four-dimensional spacetime. General relativity described gravity as curvature of this spacetime, warped by masses. Quantum mechanics destroyed the notion of determinism in particle physics. Physicists then painstakingly proceeded to put together a relativistic, renormalised quantum field theory of electromagnetism. By this time, the existence of the strong and weak nuclear interactions was recognised. The weak interaction was shown to be part of the electroweak force, along with electromagnetism. This and the strong force were both formulated as quantum field theories.

Thus electricity was unified with magnetism in the 1860s, and then with the weak interaction in the 1960s. In the 1970s, physicists explored a further possible unification with the strong force. The natural next question was whether a final unification with gravity was possible.

\medskip

Unification of gravity with another force had been attempted before - most famously by Kaluza and Klein in the 1920s, before the nuclear interactions were understood, and before renormalized, relativistic quantum field theories had been developed. Kaluza\cite{Kaluza} added a fifth coordinate to spacetime with a `cylinder condition'.  He showed that Einstein's field equations in five dimensions contained the usual four dimensional ones but also the Maxwell equations, at least within the limit of low velocities and a four-dimensional metric close to the Minkowski one. The geodesic equation also contained a Lorentz force term. (However, the correspondence failed, for example, for rapidly-moving electrons.) The action in Klein's modification of this theory\cite{Klein} was a five-dimensional generalisation of the Einstein-Hilbert action, which broke into the four-dimensional Einstein-Hilbert action and the action for the electromagnetic field. 

\medskip

But the unification of \emph{all} of the interactions will need to take into account the modern descriptions of the electroweak and strong forces. A natural place to start from is the features of these descriptions that have an analogue in classical field theory. These forces are understood in terms of gauge fields associated with local symmetries. In 1955, Utiyama\cite{Utiyama} showed that gravity could be described in a similar way. There was an erroneous assumption in his analysis, as explained below, but the basic principles are correct.

This has led many researchers to try forcing gravity into a formulation as a quantum field theory. However, the case for this way of proceeding seems far from convincing. General relativity (GR) and the quantum field theories (QFTs) of the standard model are equally successful theories within their own realms of validity, in which they have (independently) superseded the classical field theories of the preceding centuries. Moreover, the beauty and simplifying nature of the central principle of GR, that the force of gravity is simply how we perceive the curvature of spacetime, needs to be respected and developed, rather than treated as an effective theory which is valid only within limits.

The Kaluza-Klein theories of the late 1970s and early 1980s also proceeded in what could be seen in hindsight as a slightly unnatural direction, assuming the symmetry transformations of the Standard Model to act directly on the compact space formed by the extra dimensions. The compact space could itself be a group manifold or a coset space of a gauge group\cite{Schx2,SS2} or something more intricate\cite{Witten}.

\subsection{The geometry of pseudo-orthogonal and unitary symmetries and the decompactification limit}

A more thoughtful and elegant direction to proceed in is to look more deeply into the symmetries underlying the fundamental forces. The electroweak and strong forces are associated with local unitary symmetries. GR, on the other hand, is constructed to be covariant under changes of coordinate system. These induce matrix transformations on the tangent spaces. The general linear group to which these matrices belong has a pseudo-orthogonal subgroup which preserves the frame metric on these real vector spaces\cite{TSS}. (This is where Utiyama's error lay - he assumed that frame `rotations' and changes of coordinate system were independent.) In the same way, unitary transformations preserve the orthonormality of a basis on a complex vector space. However, complex vectors (such as a quark multiplet) can be combined into real vectors (such as a vector meson multiplet). The induced action of the unitary transformations on these compound real vectors is an orthogonal action, which we take to correspond to the frame-rotating transformations on the space of the additional dimensions.

This allows us to consider a de-compactification limit in which all the spatial dimensions are identical and all the fields are multiplets of the higher-dimensional general linear symmetry. We could, if desired, then construct field equations and an action which are fully covariant on this higher-dimensional spacetime.

\subsection{Product spacetimes}

We would then want to use some procedure to transition from this system to  one with a compact subspace - this is considered in Section \ref{BSS} immediately below. From existing Kaluza-Klein theories, we might expect the resulting spacetime to be some form of product space, where the compact subspace is one of the factor spaces. (Note that as it results from compactifying an isotropic pseudo-Riemannian spacetime, we would expect each of the factor spaces to be pseudo-Riemannian or Riemannian.) On such a spacetime, we might further expect the higher-dimensional general linear symmetry and its pseudo-orthogonal subgroup to be non-linearly realised. In this paper, we shall show that on a product spacetime, these symmetries are indeed non-linearly realised. 

Now, consider the vacuum of deep space from the eyes of a 19th Century physicist. It would be a space devoid of mass and charges, but also devoid of gravitational and electromagnetic fields. For the product spacetime, we will see that the equivalent of this `classical vacuum' is a Cartesian product of Minkowski spacetime and an Einstein manifold (or two-dimensional manifold). On such a spacetime, we can define coordinate systems $y^I$ which respect the factor manifolds: one subset of the coordinates, $y^\mu$, parametrises the familiar four-dimensional spacetime and the remainder, $y^X$, parametrise the other factor space - henceforth referred to as `the additional factor space'. (This is covered in more detail in Section \ref{Cartesian}. An overview of the structure of the paper as a whole is given in Section \ref{Structure}.)

In these coordinates, the higher-dimensional general linear symmetries are non-linearly realised. Only those which are induced by a transformation to \emph{another} such set of coordinates, $\{y'^\mu, y'^X\}$, are linearly realised. Non-linear realisations of symmetries allow us to decompose multiplets into those which transform according to representations of the linearly realised subgroup. (This is covered in Section \ref{tensor-decomp}, describing how the method of non-linear realisations\cite{CWZ} is applied to tensor fields.) For example, a vector multiplet $V^X$ decomposes into $V^\mu$ and $V^X$, which transform as vectors of the factor spaces. Higher-rank tensors also decompose into representations of the general linear groups on the factor spaces.

We can then consider various deviations from this `classical vacuum'. For example, we could have curvature on the familiar four-dimensional spacetime. Another possibility is not to have a \emph{Cartesian} product of the two factor spaces, but a more general product (such as a warped product or multiply warped product - see Section \ref{GPM} for a definition of a general product).

Both of these, we shall see, have clear physical interpretations. These can be summarised as follows. A coordinate system of the form $\{y^\mu, y^X\}$ can still be defined on these manifolds. In such a coordinate system, the Levi-Civita connection\footnote{In this paper, the Levi-Civita connection is denoted by a Gamma with a circle over it: $\mathring{\Gamma}$. This is to match up with the previous paper by the author\cite{TSS} and other works on teleparallel gravity, in which it is important to distinguish between this connections and others, such as the Weitzenb\"{o}ck connection. Now, the Riemann tensor is the field strength of the Levi-Civita connection (whereas the Weitzenb\"{o}ck connection, for example, has no field strength); we therefore take the approach that no circle is needed for the Riemann tensor or Ricci tensor, as no such distinction needs to be made.} used on the four-dimensional spacetime will have only indices associated with that spacetime: $\mathring{\Gamma}_{\mu \nu}{}^\rho$. Its field strength is a set of Riemann tensor components with only those indices, $R^\mu{}_{\nu \rho \lambda}$. These describe gravity. (There will also be a part of the Levi-Civita connection which has only indices associated with the additional factor space: $\mathring{\Gamma}_{XY}{}^Z$. Its field strength, $R^W{}_{XYZ}$, describes the curvature of that factor space. These will be the only components for a Cartesian product space.)

However, for the more general product of the two factor spaces, there are Levi-Civita connection components which have a mix of the indices, such as $\mathring{\Gamma}_{\mu X}{}^Y$, corresponding to geodesics which do not remain entirely in the factor spaces. We will see (in Section \ref{Gauge}) that the spin connections associated with these transform in the correct way to identify them as gauge fields of the orthogonal symmetry associated with the additional factor space. We shall show that for product spaces satisfying a higher-dimensional version of Kaluza's `cylinder condition', the covariant derivatives of vectors take exactly the right form for coupling to gravity and internal symmetries.  We will illustrate this for U(1) and SU(2). 

The field strengths for these gauge fields will turn out to be components of the Riemann tensor in a frame basis on the additional factor space, but a coordinate basis on the four-dimensional spacetime. These components do not contribute to the Ricci curvature of either the additional factor space or the four-dimensional spacetime in the $\{y^\mu, y^X\}$ coordinates.

\subsection{Breaking spacetime symmetries \label{BSS}}

For compactification models, we are then left with the question of what procedure to use to make the transition from the de-compactified space with the unbroken symmetries to the product space in which they are non-linearly realised.

For internal symmetries, the usual answer is to use a spontaneous symmetry breaking (SSB) mechanism. The general procedure is as follows\cite{SS1}. Construct a Lagrangian out of representations and, if appropriate, gauge fields of the high-energy symmetry group $J$. Include a potential constructed out of a Lorentz scalar multiplet of $J$ for which the minima are all invariant under the low-energy symmetry subgroup $G$. The vacuum manifold is then diffeomorphic to $J/G$. The Lagrangian may then be rewritten in terms of representations and, if appropriate, gauge fields of $G$. The solutions of the field equations are therefore dynamically constrained to have the $G$-symmetry.

Many of the early papers on spontaneous compactification use this approach\cite{CS,CHPS,Luciani} However, in these, the symmetry that is being broken is not a symmetry of the full higher-dimensional spacetime - it is a group action on the space of the additional dimensions only. 

In general relativity, on the other hand, one generally uses the same field equations in all situations. Solutions are then selected by hand with the desired symmetry properties, by making an ansatz such that the Einstein and energy-momentum density tensors have these properties. (This is akin to choosing the initial conditions in an initial value problem.) The papers on spontaneous compactification by Volkov, Tkach and Sorokin use this approach\cite{VT82,VST83,VST84} - and they describe the appearance of gauge fields of an orthogonal symmetry, just as we have in the models introduced in this paper.

To determine the most appropriate procedure for the compactification models (Kaluza-Klein theories) described in this paper, we focus on the way the operator forms of symmetric tensors - which we refer to as `index-aligned tensors' - transform under $GL(N, \mathbb{R})$. These tensors take values in $gl(N, \mathbb{R})$ and the group action on them is by conjugation.

This is highly reminiscent of the action of $SU(N)$ on its own Lie algebra. For $SU(N)$, this partitions the algebra into orbits\cite{MR,O'R2}. All matrices in the algebra with the same eigenvalues lie in the same orbit. The eigenvalues of a matrix are invariants under the group action and can be expressed in terms of the traces of the powers of the matrix, which are also invariants. Orbits can be collected into strata according to the multiplicities of their eigenvalues and these multiplicities determine the symmetry breaking pattern - the corresponding stabilisers are the unbroken subgroups. 

The breaking pattern is therefore determined by the traces of the powers - the `algebraic invariants'. This means that to get a specific breaking pattern, one constructs a potential which is minimised for a particular choice of algebraic invariants (or choice of relations between them). (We review the existing body of theory on all of these issues for internal symmetries in Section \ref{NLRSSB}. We also provide a brief account of how it is applied to spacetime symmetries in that section.)

We therefore look at whether this approach can be adapted for $GL(N, \mathbb{R})$. We find that the action of $GL(N, \mathbb{R})$ does partition index-aligned tensors into orbits. However, while every tensor in a given orbit has the same eigenvalues, for a spacetime with indefinite metric, not every tensor with the same eigenvalues lies in the same orbit. Every \emph{diagonalisable} tensor with a given set of eigenvalues lies in the same orbit, but for spacetimes with indefinite metric, not every index-aligned tensor is diagonalisable. (This is covered in Section \ref{diag-non}.)

In the absence of an algebraic way of ensuring an index-aligned tensor is diagonalisable across a chart, we are forced to abandon the idea of using a Goldstone/Higgs-type potential to break symmetry dynamically.

However, we are able to show (in Section \ref{Constraints}) that if any real index-aligned tensor is diagonalisable across a chart with the same multiplicities of eigenvalues at each point, this is a necessary and sufficient condition for the spacetime to be isometric to a product manifold across that chart. Furthermore, the dimensionalities of the factor spaces are equal to the multiplicities of its eigenvalues. This allows the decomposition of all tensor fields across the chart.

It does not imply, however, that all index-aligned tensors are simultaneously diagonalisable. For example, the diagonalisable tensor may relate to a particular form of matter in the system. It may be that other index-aligned tensors relating to other forms of the matter in the system do not share this property. 

Note, though, that if Einstein's field equations are valid for the system, then the matter content of the system as a whole is described by an energy-momentum density tensor, which has the same multiplicities of eigenvalues as the Einstein tensor, Ricci tensor and traceless Ricci tensor. 

This begs the question of what happens when these rank-two curvature tensors are diagonalisable across a chart, with the same multiplicities of eigenvalues at each point. We find in this case (in Section \ref{PRIs}) that the spacetime coincides locally with a product of Einstein manifolds and two-dimensional manifolds. The algebraic invariants are `pure Ricci invariants' - higher-dimensional generalisations of those used by Carminati and McLenaghan\cite{CM2} and Narlikar and Karmarkar (as summarised and rectified by Harvey\cite{Harvey}). We provide an example - the pure Ricci invariants corresponding to the classical vacuum for our model in six dimensions.

The algebraic invariants therefore classify the product spacetime solutions of the field equations, for any theory in which a real, symmetric geometric tensor is proportional to a tensor describing the distribution of matter in the system. This makes sense - we know that Einstein's field equations have many solutions, corresponding to different matter content and different geometries. In any theory in which, in Wheeler's famous phrase `Spacetime tells matter how to move; matter tells spacetime how to curve', it is natural that the symmetries of one are reflected in the symmetries of the other.

\subsection{Structure of this paper \label{Structure}}

The following summarises the layout of this paper, and the main results of each section (in italics). Note that in the early sections of the paper, particularly from Section \ref{Simplest} to Section \ref{GPM}, many of the main results are trivial or straightforward to derive. However, they are worth expressing in these forms, (which rarely, if ever, appear in the literature), as they form the basis for the analysis in the later sections.

\smallskip

The paper is in two parts. 

\smallskip

The first part of the paper, from Section \ref{Tensor} to Section \ref{Gauge}, analyses the geometric tensors and connections of product manifolds and their factor spaces. This allows us to construct Kaluza-Klein theories on these spaces.

In Section \ref{Tensor}, we consider changes of coordinate system on a pseudo-Riemannian manifold, and describe the action these induce on tangent spaces and their duals, and on outer products of them. We use a formalism based on \cite{TSS} and draw out the group theory aspects of changes of bases. We examine in detail the action on the operator forms of the metric (that is, the Kronecker delta) and other symmetric tensors.

In Section \ref{Simplest}, we apply the analysis to the Ricci tensor for three types of manifold: Ricci-flat, Einstein and two-dimensional. \emph{We show that the operator form of the Ricci tensor for any of these manifolds is diagonal in every coordinate system, with all eigenvalues equal. The same is true of the Einstein and traceless Ricci tensors - and the total energy-momentum density tensor if the Einstein field equations hold.}

In Section \ref{Cartesian}, we look at Cartesian product manifolds. This allows us to identify the classical vacuum of our Kaluza-Klein theories. We define coordinate systems which do and do not respect the factor manifolds. The latter type of coordinate system is illustrated with that implicitly used in the Klein metric. We also introduce the concept of the stabiliser of the Ricci tensor. \emph{In this section, we show that:
	\begin{itemize}
		\item in any coordinate system which respects the factor manifolds, the operator form of the Ricci tensor for any Cartesian product of Einstein manifolds and two-dimensional manifolds is diagonal, with each of the eigenvalues associated with a given factor space being equal;
		\item in any coordinate system which respects the factor manifolds, for \emph{any} Cartesian product manifold, the metric and the Ricci tensor in both its operator form and completely covariant form are block diagonal. The Levi-Civita connection and Riemann tensor also decompose solely into parts for each factor space.
\end{itemize} }

In Section \ref{GPM}, we define the more general class of product manifolds. We note that by definition, in any coordinate system which respects the factor spaces, the metric takes block diagonal form, just as for Cartesian product manifolds. We illustrate this with a tube of varying radius. For the most general product manifolds, the Riemann tensor may have components which are absent for Cartesian product manifolds.  

\emph{But for any product of Einstein manifolds and two-dimensional manifolds, we again show that in any coordinate system which respects the factor manifolds, the operator form of the Ricci tensor is diagonal, with each of the eigenvalues associated with a given factor space being equal.} 

In Section \ref{Subgroups}, we look at coset space decompositions of changes of basis on the tangent space. This draws on theory developed for the non-linear realisations of internal symmetries. We show that a general change of basis can be decomposed using its pseudo-orthogonal subgroup. This can then be decomposed further into its connected subgroup and a discrete group. 

Alternatively, we can decompose a general change of basis using the subgroup of basis changes on the factor manifolds. This leads us to the follow results: 
\begin{itemize}
	\item \emph{let $J$ be the group of all changes of basis on an $N$-dimensional manifold. Let $G$ be isomorphic to a subgroup $GL(n_1, \mathbb{R}) \otimes GL(n_2, \mathbb{R}) \otimes \ldots$, such that $\sum n_i = N$. Then if a coset space representative $L \in J/G$ can be consistently be defined across a neighbourhood, the manifold is isometric to a product manifold across that neighbourhood;}
	\item \emph{on such a space, all tensors decompose into multiplets of the factor groups of $G$.}
\end{itemize}

In Section \ref{Gauge}, we start by looking more closely at components of the Levi-Civita connection and Riemann tensor which have indices relating to both factor spaces. For Cartesian product manifolds, we find that the Levi-Civita connection only has such components in coordinate systems which do not respect the factor manifolds. Even in these, the Riemann tensor does not have such components - there is no additional intrinsic curvature; the new components of the Levi-Civita connection can be `gauged away' with a change of coordinate system. For more general product manifolds, on the other hand, these components of the connection exist even in the coordinate systems which respect the factor manifolds. This reflects intrinsic curvature. We also view these issues in terms of holonomy.

For a general (that is, non-Cartesian) product manifold which nonetheless satisfies a generalised form of Kaluza's `cylinder condition', the additional connection components transform as four-vectors and as gauge potentials of the general linear group on the compact space. We then point out that such a connection has an associated Lorentz (spin) connection, and show that:
\emph{
	\begin{itemize}
		\item this Lorentz connection transforms as a four vector and as a gauge potential for the orthogonal transformations on the compact space;
		\item its field strength may be written in terms of the additional components of the Riemann tensor;
		\item with two extra dimensions, a six-vector decomposes into a four-vector and a two-vector. The latter can be expressed as a complex field and couples to the gauge field in exactly the right way to identify the gauge field as the electromagnetic potential;
		\item with three extra dimensions, a seven-vector decomposes into a four-vector and a three-vector. The latter couples to the gauge field in exactly the right way for the coupling of an $SU(2)$ gauge field to an $SU(2)$ triplet.
\end{itemize} }

We close the section by pointing out that for a product of two manifolds, we can separate the components of the Riemann tensor into four classes. The components which reflect the field strength of the gauge fields do not contribute to the Ricci tensor, in the coordinate systems which respect the factor space.

\smallskip

The second part of the paper looks at how index-aligned tensors transform under the action of the higher-dimensional general linear group. We use this to study the transition in models of compactification from the decompactified higher-dimensional spacetime to the product manifold. It draws heavily on the results from the first part. (This allows fuller expressions of theorems that might be useful to geometers working outside the research field of Kaluza-Klein theories.)

We start in Section \ref{NLRSSB} by looking at the analogous transition in the breaking or non-linear realisation of internal symmetries. We focus particularly on the case in which the Goldstone/Higgs field space is diffeomorphic to the Lie algebra of the unitary unbroken internal group. We also provide a brief account of how these concepts are applied to spacetime symmetries in the existing literature.

In Section \ref{Subspaces}, we see that rank-two tensors in their operator form take values in the Lie algebra of the general linear group. \emph{Index-aligned tensors take values in a subspace of this. The action of the group on this subspace is by conjugation}, just as it is in the case of a special unitary group acting on its Lie algebra.

In Section \ref{diag-non}, we note that this action preserves the eigenvalues of the tensors and partitions the set of all possible values into orbits. We consider whether the action may used to diagonalise an index-aligned tensor. For a space of indefinite metric, an index-aligned tensor has eigenvalues which may be real or complex:
\begin{itemize}
	\item \emph{if any of them are complex, it cannot be diagonalised by a change of coordinates;
		\item if they are all real and distinct, it can always be diagonalised by a change of coordinates (or, indeed, if the space has positive definite metric} - both of these are proved in the Appendix);
	\item \emph{if they are all real and some are repeated, it may or may not be diagonalisable using a change of coordinates - one needs to consider the eigenspaces.} 
\end{itemize}
Indeed, it is possible for two tensors to have the same set of eigenvalues, but one is diagonalisable and the other is not.

Section \ref{Stab} looks at Cartan subspaces of the Lie algebra. We find that 
\emph{
	\begin{itemize}
		\item the group action maps one Cartan subalgebra to another;
		\item for $GL(N,\mathbb{R})$, each can be spanned by a set of $N$ linearly independent, commuting matrices - for example, the powers of an index-aligned tensor with distinct eigenvalues (or by a set of projection operators, as shown in the Appendix);
		\item for a diagonalisable tensor, its stabiliser group under a change of coordinates depends on the multiplicities of its eigenvalues and is always a product of general linear groups.
\end{itemize} }

In Section \ref{Constraints}, we put together the results of several previous sections, to find that \emph{if there exists an index-aligned tensor field which may be diagonalised across a coordinate neighbourhood with the same multiplicities of eigenvalues at each point:
	\begin{itemize}
		\item then it has the same stabiliser group across that neighbourhood (up to isomorphism);
		\item the group of basis changes may be decomposed with respect to this stabiliser group, and the tensor field can then be diagonalised across the neighbourhood using just a representative of the corresponding coset space;
		\item the manifold must coincide across that neighbourhood with some form of product manifold, for which the factor spaces have the same dimensionalities as the multiplicities of the tensor field's eigenvalues;
		\item in any coordinate system which respects the factor spaces of such a manifold, a) this tensor takes diagonal form, b) all tensor fields may be decomposed into tensors of the individual factor spaces, and c) the metric takes block diagonal form;
		\item the stabiliser group depends on the multiplicities of the eigenvalues. These eigenvalues are completely determined by the traces of the powers of the tensor field, so they provide a coordinate-independent way of specifying the gauge fields (at least, up to changes of signature);
		\item however, there does not appear to be any obvious way of requiring such a tensor to be diagonalisable using these algebraic invariants;
		\item consequently, it looks unlikely that a Goldstone/Higgs-type potential can be used to provide compactification to the types of Kaluza-Klein theories described in this paper.
\end{itemize} }
It is worth remarking here that an $N$-dimensional pseudo-Riemannian manifold coinciding across a neighbourhood with a product manifold is not unusual. Indeed, those manifolds which do not have such a decomposition are the exceptional ones. Any manifold for which the metric can be diagonalised across a neighbourhood is, by definition, isometric to a product of one-dimensional manifolds across this neighbourhood. Most metrics commonly studied by physicists are of this type. (An example of one which cannot be diagonalised across a neighbourhood is the Kerr metric, but even this takes a block diagonal form.)

What is of interest to us here is when a higher-dimensional geometry may factorise in such a way as to support the kind of Kaluza-Klein theory described in the first part of the paper.

\smallskip

In Section \ref{PRIs} we apply these results to the case where the tensor field is the Ricci tensor itself. This allows us to prove the converse of our theorem from Section \ref{GPM}. \emph{Whenever the Ricci tensor is diagonalisable across a neighbourhood with the same multiplicities of eigenvalues at each point,
	\begin{itemize}
		\item the manifold \emph{must necessarily} coincide across that neighbourhood with a Cartesian product of Einstein manifolds and two-dimensional manifolds;
		\item in any coordinate system which respects the factor spaces of such a manifold, a) the operator form of the Ricci tensor is diagonal and b) the metric and completely covariant Ricci tensor are block diagonal;
		\item the eigenvalues of the operator form of the Ricci tensor may be determined by the traces of its powers, known as the pure Ricci invariants.
\end{itemize} }

Finally, in Section \ref{Concs}, we summarise the main results. We also briefly discuss three other issues:
\begin{itemize}
	\item what happens to higher-dimensional translations when the spacetime is compactified and how this takes advantage of a loophole in O'Raifeartaigh's no-go theorem;
	\item the need for an action, or at least field equations;
	\item the inclusion of spinors and the importance of Clifford algebra structures to the inclusion of higher-dimensional unitary symmetries.
\end{itemize}
All three of these will be the focus of future research.

\section*{Part I: The geometry of product manifolds \label{Geom-PMs}}

\section{Tensor spaces and group actions on them \label{Tensor}}

The basic symmetries in our theory are symmetries of general linear groups and their orthogonal or pseudo-orthogonal subgroups, so we devote this section to describing their transformations. Their actions on rank-two symmetric tensors in their mixed (operator) form will be particularly important to our study.

\subsection{The tangent space}

Consider a pseudo-Riemannian manifold $\mathcal{M}$ with $t$ timelike dimensions and $s$ spacelike dimensions, on which there is a coordinate neighbourhood $\Omega$ with coordinates $u^I$. The vectors tangent to the curves of increasing $u^0, u^1, u^2, \ldots$ at a point $A$ form a basis for the tangent space $T_A \mathcal{M}$, denoted $\mathbf{e}_I |_A$ - the ``coordinate basis'' for $u^I$. The value of a vector field at $A$ may then be written as a linear sum of this coordinate basis
\begin{equation}
	\mathbf{V} |_A \in T_A \mathcal{M} = V_{(u)}^M |_A \mathbf{e}_M |_A 
\end{equation}
Under a change of coordinates $u^I \rightarrow u'^I$, this changes according to the rule
\begin{equation}
	V_{(u)}^I |_A \mathbf{e}_I |_A = V_{(u)}^I |_A \left. \frac{\partial u'^J}{\partial u^I} \right|_A \mathbf{e'}_J |_A
\end{equation}
We can see this as a transformation of either the basis:
\begin{equation} \label{e-to-e'}
	\mathbf{e}_I |_A = \left. \frac{\partial u'^J}{\partial u^I} \right|_A \mathbf{e'}_J |_A
\end{equation}
or the components:
\begin{equation} \label{V-to-V'}
	V_{(u')}^J |_A = V_{(u)}^I |_A \left. \frac{\partial u'^J}{\partial u^I} \right|_A
\end{equation}

The Jacobian matrices for transforming between bases at $A$ are invertible $N \times N$ real matrices, where $N = t + s$. The set of all such matrices thus forms a group $J_A$ which is isomorphic to $GL(N,\mathbb{R})$. 

Being a pseudo-Riemannian manifold, a symmetric inner product is defined on each tangent space:
\begin{equation} \label{inner}
	(\mathbf{V}, \mathbf{W})_A = (\mathbf{W}, \mathbf{V})_A \in \mathbb{R} \, .
\end{equation}

\subsection{The dual space and rank-two tensors \label{Dual}}

$T_A \mathcal{M}$ has a dual space $T^*_A \mathcal{M}$. One possible basis for this space is the dual of $\mathbf{e}_I |_A$, denoted $\mathbf{e}^I |_A$. We can then take products of the tangent space and its dual to construct higher-order tensor spaces. Outer products of $\mathbf{e}_I |_A$ and $\mathbf{e}^I |_A$ then form bases for these spaces. We will be particularly interested in rank-two tensor spaces. There are four of these, for which we can construct the following bases:
\begin{eqnarray}
	\mathbf{e}_I |_A \otimes \mathbf{e}_J |_A &\in& T_A \mathcal{M} \otimes T_A \mathcal{M} \\
	\mathbf{e}^I |_A \otimes \mathbf{e}^J |_A &\in& T^*_A \mathcal{M} \otimes T^*_A \mathcal{M} \\
	\mathbf{e}_I |_A \otimes \mathbf{e}^J |_A &\in& T_A \mathcal{M} \otimes T^*_A \mathcal{M} \\
	\mathbf{e}^I |_A \otimes \mathbf{e}_J |_A &\in& T^*_A \mathcal{M} \otimes T_A \mathcal{M}
\end{eqnarray}

The image of the map (\ref{inner}) on the coordinate basis is the metric in those coordinates at $A$:
\begin{equation}
	\mathrm{g}_{IJ} |_A = (\mathbf{e}_I, \mathbf{e}_J)_A \, .
\end{equation}
This itself may be seen as the set of components of a tensor:
\begin{equation}
	\mathbf{g} |_A = \mathrm{g}_{IJ} |_A \; \mathbf{e}^I |_A \otimes \mathbf{e}^J |_A \in T^*_A \mathcal{M} \otimes T^*_A \mathcal{M} 
\end{equation}

We can always define a set of Riemann normal coordinates $x^I$ whose basis $\hat{\mathbf{n}}_I$ is pseudo-orthonormal at $A$ \cite{TSS}:
\begin{equation}
	\eta_{IJ} |_A = (\hat{\mathbf{n}}_I, \hat{\mathbf{n}}_J)_A \, ,
\end{equation}
where $\eta_{IJ} = diag(1,-1,-1,-1)$ is the Minkowski metric.

We can also define an inner product on the dual space, which provides us with the contravariant metrics $\eta^{IJ} |_A$ and $\mathrm{g}^{IJ} |_A$. These are the inverses of the covariant metrics:
\begin{equation} \label{eta-inv}
	\eta_{IJ} |_A \, \eta^{JK} |_A = \delta_I^K
\end{equation}
and
\begin{equation} \label{g-inv}
	\mathrm{g}_{IJ} |_A \, \mathrm{g}^{JK} |_A = \delta_I^K
\end{equation}
They can be used to raise and lower the indices of tensors. For example, they 
may be used to raise one index of the completely covariant form of any rank-two tensor:
\begin{equation} \label{Xmixedx}
	X_{(x)}^I{}_K |_A = \eta^{IJ} \, X^{(x)}_{JK} |_A
\end{equation}
and 
\begin{equation} \label{Xmixedu}
	X_{(u)}^I{}_K |_A = \mathrm{g}^{IJ} |_A \, X^{(u)}_{JK} |_A
\end{equation}
where $X^I{}_K |_A$ are the components of an element of $T_A \mathcal{M} \otimes T^*_A \mathcal{M}$:
\begin{equation}
	X_{(x)}^I{}_K |_A \, \hat{\mathbf{n}}_I |_A \otimes \hat{\mathbf{n}}^K |_A = X_{(u)}^I{}_K |_A \, \mathbf{e}_I |_A \otimes \mathbf{e}^K |_A \in T_A \mathcal{M} \otimes T^*_A \mathcal{M}
\end{equation}

Observe that this mixed form of a rank-two tensor may be used to map a vector to another vector:
\begin{equation}
	X^I{}_K |_A: V^K |_A \mapsto V'^I |_A = X^I{}_K |_A V^K |_A
\end{equation}
and consequently this mixed form of a tensor is known as the operator form of the tensor.

\subsection{Fields on a chart}

If a coordinate system $u^I$ is valid over a coordinate neighbourhood $\Omega$, the pair $(\Omega, u^I)$ is known as a chart\cite{Nakahara}. We may define tensor fields such as $X_{JK} (u)$. The metric in $u$-coordinates is one such field, $\mathrm{g}_{IJ} (u)$. Its contravariant form $\mathrm{g}^{IJ} (u)$ is also a field. This allows us to form contravariant and mixed forms of our tensor fields.

If we have two coordinate systems $u^I$ and $u'^I$ defined over a coordinate neighbourhood, then the matrix which relates the bases for the two coordinate systems across this neighbourhood is also promoted to a field:
\begin{equation}
	\mathbf{e}_I = \frac{\partial u'^J}{\partial u^I} \mathbf{e'}_J
\end{equation}
These belong to a group of local transformations which we shall call $J$.

In this transition to looking at coordinate neighbourhoods, we must remember that on a curved spacetime, the metric for the Riemann normal coordinates $x^I$ cannot be pseudo-orthonormal beyond a geodesic passing through $A$ \cite{TSS}.

\subsection{Action of $J$ on rank-two tensors}

Under a change of coordinates across a neighbourhood, a tensor field with two covariant indices transforms according to
\begin{equation}
	X^{(u')}_{IK} = \frac{\partial u^J}{\partial u'^I} \frac{\partial u^L}{\partial u'^K} X^{(u)}_{JL}
	= j_I{}^J j_K{}^L X^{(u)}_{JL}
\end{equation}
where
\begin{equation}
	j_K{}^L = \frac{\partial u^L}{\partial u'^K} \in J
\end{equation}
For example, the metric transforms in this way:
\begin{equation} \label{metric-map}
	\mathrm{g}^{(u')}_{IK} = j_I{}^J j_K{}^L \mathrm{g}^{(u)}_{JL}
\end{equation}
Conversely, a tensor with two contravariant indices transforms according to 
\begin{equation}
	X_{(u')}^{IK} = \frac{\partial u'^I}{\partial u^J} \frac{\partial u'^K}{\partial u^L} X_{(u)}^{JL}
	= (j^{-1})^I{}_J (j^{-1})^K{}_L X_{(u)}^{JL}
\end{equation}
and a rank-two tensor in mixed form transforms by conjugation under $J$:
\begin{equation} \label{j-action}
	X'_K{}^J = \frac{\partial u^L}{\partial u'^K} X_L{}^I \frac{\partial u'^J}{\partial u^I} = (j X j^{-1})_K{}^J
\end{equation}

It is well known that $\eta_{IJ}$ is invariant under $I \subset J$ isomorphic to local $O(t,s)$. This group has $\frac{1}{2} N (N-1)$ dimensions\cite{Einstein1928} and acts on the basis $\mathbf{e}_I$. Denoting an arbitrary element of it $i$, the invariance condition is then
\begin{equation}
	i: \eta_{IJ} \mapsto \eta'_{IJ} = i_I{}^K \, i_J{}^L \, \eta_{KL} = \eta_{IJ}
\end{equation}
Contracting both sides with $(i^{-1})_M{}^I \, \eta^{JN}$, we get
\begin{eqnarray}
	\delta^K_M \, \eta^{JN} \, i_J{}^L \, \eta_{KL} &=& (i^{-1})_M{}^I \delta_I^N \\
	\Rightarrow i^N{}_M &=& (i^{-1})_M{}^N
\end{eqnarray}
This is the condition for the group to be pseudo-orthonormal. Note that if we consider the action under $i^{-1}$ instead, we get
\begin{equation}
	(i^{-1})^N{}_M = i_M{}^N
\end{equation}

\subsection{The operator forms of the metric and other symmetric tensors}

Observe that in (\ref{eta-inv}) we are effectively raising one index of $\eta_{IJ}$, while in (\ref{g-inv}) we are raising one index of $\mathrm{g}_{IJ}$. $\eta_{IJ}$ contains information about the signature of $\mathcal{M}$, while $\mathrm{g}_{IJ}$ carries $\frac{1}{2} N (N+1)$ degrees of freedom\cite{TSS,Einstein1928}. The mixed form of the metric, $\delta_I^J$, contains neither of these - it is completely fixed. It is the same in all coordinate systems, because it is invariant under (\ref{j-action}). That is, under the action (\ref{j-action}) it is stabilised by the whole of $J$. We shall explore this issue of the stabiliser under (\ref{j-action}) later in this paper.

Note also that the indices of the Kronecker delta are aligned. This also conveys subtle but important information. In general, a rank-two tensor field has two distinct mixed forms, (\ref{Xmixedu}) and
\begin{equation} \label{Xmixedu2}
	X^{(u)}_K{}^I = X^{(u)}_{KJ} \, \mathrm{g}^{IJ}
\end{equation}
However, for a tensor for which the completely covariant form is symmetric:
\begin{equation} \label{symm}
	X_{IJ} = X_{JI}
\end{equation}
we can see by comparing (\ref{Xmixedu}) and (\ref{Xmixedu2}) that these mixed forms are equal:
\begin{equation} \label{IswapK}
	X_K{}^I = X^I{}_K
\end{equation}
which allows us to align the indices. Note that this is equating a matrix of components in $T^*_A \mathcal{M} \otimes T_A \mathcal{M}$ with a matrix of components in $T_A \mathcal{M} \otimes T^*_A \mathcal{M}$, for each space $A$ tangent to the neighbourhood. It is not the same thing as saying that the matrix is symmetric. The matrix may be written
\begin{equation}
	\left( \begin{array}{cccccc}
		X^0_0 & X^0_1 & X^0_2 & X^0_3 & X^0_5 \ldots & X^0_{N} \\
		X^1_0 & X^1_1 & & & & \\
		X^2_0 & & \ddots  & & & \\
		\vdots & & & & & \\
		X^{N}_0 & & & & & X^{N}_{N}
	\end{array} \right)
\end{equation}
(we follow the common, somewhat eccentric, convention of using 0 to denote the time coordinate, 1, 2 and 3 to denote the three observed spatial dimensions and the numbers starting from 5 to denote additional spatial dimensions). This matrix is symmetric if 
\begin{equation}
	X^I_K = X^K_I \hspace{5mm} \forall \, I,K
\end{equation}

We shall refer to a tensor which satisfies the property (\ref{IswapK}) as an `index-aligned tensor' (rather than the more long-winded `operator form of a symmetric tensor'). 

We shall explore this further from Section \ref{Subspaces} onwards. For now, we shall note that the action of $J$ on $X^J{}_I$ is easily shown to be
\begin{equation}
	j \in J: X^J{}_I \mapsto X'^J{}_I = j_I{}^L X^K{}_L (j^{-1})_K{}^J
\end{equation}
so if $X^K{}_L$ has the property (\ref{IswapK}), then this property is preserved under changes of coordinate:
\begin{equation} \label{symm-rep}
	j \in J: X^J{}_I \mapsto X'^J{}_I = X'_I{}^J
\end{equation}
Now, any completely covariant rank-two tensor may be decomposed into symmetric and anti-symmetric parts. If $\bar{X}_{IJ}$ is an anti-symmetric tensor, then
\begin{equation} \label{antisymm}
	\bar{X}^J{}_I = \mathrm{g}^{JK} \bar{X}_{KI} = - \mathrm{g}^{JK} \bar{X}_{IK} = - \bar{X}_I{}^J
\end{equation}
We thus see that (\ref{symm-rep}) tells us that the action (\ref{j-action}) transforms an index-aligned tensor into another index-aligned tensor, without mixing it with one with the property (\ref{antisymm}). That is, $X_I^J$ is a representation of $J$.

\section{The simplest manifolds: Ricci-flat, Einstein and two-dimensional \label{Simplest}}

Having understood the transformation properties of vectors and tensors, we now want to turn to the geometries of particular spacetimes. We look at the simplest of these in this section, which will be relevant to the classical vacuum of our Kaluza-Klein models. 

The key tensor in this study is the Ricci tensor, which in its operator form is an index-aligned tensor, so we can apply the analysis of the last section to it.

The Ricci tensor is symmetric in its completely covariant form, so the matrices for its mixed forms are equal and may be written $R_I^J$. It therefore forms a representation of $J$, but this is not an irreducible representation, as its trace, the Ricci scalar $R$, is a singlet under $J$. This may be subtracted to create the traceless Ricci tensor
\begin{equation}
	S_I^J = R_I^J - \frac{1}{N} R \delta_I^J
\end{equation}
which also transforms as a representation of $J$ and is used in constructing three of the Carminati-McLenaghan invariants (see below). If we replace the factor of $1/N$ with $1/2$ we get the Einstein tensor
\begin{equation}
	G_I^J = R_I^J - \frac{1}{2} R \delta_I^J
\end{equation}
which satisfies the divergenceless condition of the contracted Bianchi identity.

\medskip

The simplest spacetime is a Ricci-flat one. In a Ricci-flat manifold or Ricci-flat region of a manifold, the three tensors are zero everywhere:
\begin{equation}
	R_I^J = S_I^J = G_I^J = 0
\end{equation}
If the field equations of general relativity (without cosmological constant) apply, the field equations tell us that the Einstein tensor is proportional to the total energy-momentum density tensor, so
\begin{equation}
	T_I^J = 0
\end{equation}
This will obviously be the case for any matter-free region, even if there is Riemann curvature due to nearby matter. These equations, being tensor equations, are valid in all coordinate systems.

\medskip

The next simplest spacetime is one for which the Ricci tensor is proportional to the metric:
\begin{equation}
	R_{IJ} \propto \mathrm{g}_{IJ}
\end{equation}
By contraction, we find that this may be written
\begin{equation} \label{Ricci-Ein}
	R_{IJ} = \frac{R}{N} \mathrm{g}_{IJ}
\end{equation}
If this is valid across a coordinate neighbourhood, then substituting this into the contracted Bianchi identity in the form
\begin{equation}
	D_J R_I^J = \frac{1}{2} D_I R
\end{equation}
we rapidly find that for $N \neq 2$,
\begin{equation}
	D_I R = \partial_I R = 0
\end{equation}
so that $R$ is constant, and the manifold coincides with an Einstein manifold, across the coordinate neighbourhood. For \emph{any} two-dimensional manifold, (\ref{Ricci-Ein}) is valid, but there is no requirement for $R$ to be constant.

If the Einstein field equations hold, (\ref{Ricci-Ein}) implies that the total energy-momentum density tensor is at least locally proportional to the metric. (Note that if there is more than one type of matter in the system, this does not require the energy-momentum density tensor for each of them to be proportional to the metric, just the total for all of them.)

In the form (\ref{Ricci-Ein}), the Ricci tensor looks simple enough, but it looks even simpler in the mixed form:
\begin{equation}
	R_I^J = \frac{R}{N} \delta_I^J = \frac{1}{N} \begin{pmatrix}
		R & &  & \\
		& R & & \\
		&  & \ldots & \\
		& & & R
	\end{pmatrix}
\end{equation}
We thus see that for any Einstein manifold or two-dimensional manifold, this matrix of components is diagonal with all eigenvalues equal. Because it is proportional to the identity matrix, it is stabilised by the whole of $J$. This implies that $R_I^J$ takes this form in \emph{every} coordinate system at the specified point in spacetime. $S_I^J$ and $G_I^J$ also take this form - diagonal with all eigenvalues equal - in all coordinate systems.

\section{Cartesian product spacetimes and the Klein metric \label{Cartesian}}

We now want to see how manifolds can be assembled into the kinds of product manifolds that appear in Kaluza-Klein theories. We start with the simplest cases: Cartesian products of Einstein manifolds and/or two-dimensional manifolds.

\subsection{Results in $y$-coordinates}

Let $\mathcal{M}^1$ be any Einstein manifold (including a Ricci-flat one) or two-dimensional manifold and let $y^\mu$ be a set of arbitrary curvilinear coordinates on it. Let $\mathcal{M}^2$ also be an Einstein or two-dimensional manifold and let $y^X$ be a set of arbitrary curvilinear coordinates on it.

Then if $\mathcal{M}$ is isometric to the Cartesian product $\mathcal{M}^1 \times \mathcal{M}^2$ across the neighbourhoods on which these coordinates are valid, the combined set $y^I = \{ y^\mu, y^X \}$ is a coordinate system on $\mathcal{M}$. We can describe this as a coordinate system which respects the factor spaces. In such a coordinate system, the metric for $\mathcal{M}$ takes the form
\begin{equation} \label{metric-Cart}
	\mathrm{g}_{IJ} = \left( \begin{array}{cc}
		\mathrm{g}_{\mu \nu} (y^\rho) & 0 \\
		0 & \mathrm{g}_{XY} (y^Z)
	\end{array} \right)
\end{equation}
The condition (\ref{Ricci-Ein}) then means that the Ricci tensor takes the form
\begin{equation}
	R_{IJ} = \left( \begin{array}{cc}
		\frac{R_1}{N_1} \mathrm{g}_{\mu \nu} (y^\rho) & 0 \\
		0 & \frac{R_2}{N_2} \mathrm{g}_{XY} (y^Z)
	\end{array} \right)
\end{equation}
where $N_1$ and $N_2$ are the dimensionalities of $\mathcal{M}^1$ and $\mathcal{M}^2$ respectively and $R_1$ and $R_2$ are their Ricci scalars. In operator form, we then have
\begin{equation} \label{Ric2}
	R_I^J = \left( \begin{array}{cc}
		\frac{R_1}{N_1} \delta_\mu^\nu & 0 \\
		0 & \frac{R_2}{N_2} \delta_X^Y
	\end{array} \right)
\end{equation}
This can trivially be extended to a product of any number of such manifolds. Thus in the coordinate system $(y^\mu, y^Z, \ldots)$, the operator form of the Ricci tensor for any Cartesian product of Einstein manifolds and two-dimensional manifolds is diagonal, with each of the eigenvalues associated with a given factor space being equal. The stabiliser of such a matrix is then $GL(N_1, \mathbb{R}) \otimes GL(N_2, \mathbb{R}) \otimes \ldots \otimes GL(N_n, \mathbb{R})$. 

The case where $\mathcal{M}^1$ is a flat manifold, with one time dimension and three spatial dimensions, represents the full classical vacuum of our Kaluza-Klein models. (Far away from gravitating matter, $\mathcal{M}^1$ would be Riemann-flat; outside gravitating matter but close enough that gravitational effects are felt, it would be Ricci-flat.)

The form (\ref{metric-Cart}) of the metric also applies if $\mathcal{M}^1$ and $\mathcal{M}^2$ are not Einstein manifolds or two-dimensional manifolds. For such a Cartesian product of more general manifolds, the Ricci tensors of the factor spaces are no longer proportional to the metrics for those spaces. However, it is easy to see that the Levi-Civita connection decomposes into parts for each factor space. The Riemann tensor then decomposes in this same way, as does the contravariant metric. (We will look more closely at the physical meaning of these decompositions in Section \ref{Gauge}.)
This means that the Ricci tensor decomposes into tensors of $\mathcal{M}^1$ and $\mathcal{M}^2$ in both its totally covariant and mixed forms, with the same block diagonal form as the metric:
\begin{equation}
	R_{IJ} = \left( \begin{array}{cc}
		R_{\mu \nu} (y^\rho) & 0 \\
		0 & R_{XY} (y^Z)
	\end{array} \right)
\end{equation}
and
\begin{equation}
	R_I^J = \left( \begin{array}{cc}
		R_\mu^\nu (y^\rho) & 0 \\
		0 & R_X^Y (y^Z)
	\end{array} \right)
\end{equation}

The Jacobian matrix for a change of coordinates on $\mathcal{M}^1$ is an element of a group $G_1 \cong GL(N_1,\mathbb{R})$, where $N_1 = t_1 + s_1$ and $\cong$ denotes an isomorphism. Similarly, the Jacobian matrix for a change of coordinates on $\mathcal{M}^2$ is an element of a group $G_2 \cong GL(N_2,\mathbb{R})$, where $N_2 = t_2 + s_2$. This means that the metric and Ricci tensor retain this block diagonal form in any coordinate system which respects the factor manifolds. Again, this can be trivially extended to a Cartesian product of any number of manifolds.

\subsection{General coordinate systems and the Klein metric}

However, the most general coordinates on the product manifold are functions of both $y^\mu$ and $y^X$. In such coordinate systems, the metric does not take block diagonal form. 

We can illustrate this point with the Klein metric. Klein's theory was based on spacetime isometric to $\mathcal{M}^4 \times S^1$, where $\mathcal{M}^4$ is a curved four-dimensional spacetime. If we take $y^5$ to be the angular variable $\phi$ on the $S^1$ factor space, then the metric in $y$-coordinates is
\begin{equation}
	\mathrm{g}_{IJ} = \left( \begin{array}{cc}
		\mathrm{g}_{\mu \nu} (y^\rho) & 0 \\
		0 & r^2
	\end{array} \right)
\end{equation}
where $r$ is constant over all coordinates. We now make a change of coordinates, replacing $y^5$ with 
\begin{equation} \label{A-Klein}
	u^5 = \phi - A_\mu y^\mu
\end{equation}
- a linear sum of all five coordinates, where $A_\mu$ is the electromagnetic gauge potential. This gives us the Klein metric:
\begin{equation} \label{K-met}
	\mathrm{g}_{IJ} = \left( \begin{array}{cc}
		\mathrm{g}_{\mu \nu} + k A_\mu A_\nu & k A_\mu \\
		k A_\nu & k
	\end{array} \right)
\end{equation}
with $k = r^2$.

For a Cartesian product manifold in such a coordinate system, the mixed Ricci tensor will not in general take a block diagonal form either. 

By considering two such coordinate systems, we see that the group of all possible changes of basis is still $J \cong GL(N, \mathbb{R})$, where $N = N_1 + N_2$ (or more generally, $N = N_1 + N_2 + N_3 + \ldots$). $G_1$ and $G_2$ are mutually commuting subgroups of $J$. (We will explore this point further in Section \ref{Subgroups}.)

\section{General product manifolds - Definition and simple results \label{GPM}}

While our Cartesian product manifold exhibits gravitational effects when $\mathcal{M}_1$ is curved, we have yet to see the appearance of gauge fields. For this, we need to look at general product manifolds.

We define a general product manifold as one for which the metric in some appropriate system of coordinates ($y$-coordinates) takes the form
\begin{equation} \label{block-met}
	\mathrm{g}_{IJ} = \left( \begin{array}{cc}
		\mathrm{g}_{\mu \nu} (y^\rho, y^Z) & 0 \\
		0 & \mathrm{g}_{XY} (y^\rho, y^Z)
	\end{array} \right)
\end{equation}

\begin{figure} [ht]
	\centering
	\begin{overpic}[width=1\linewidth]{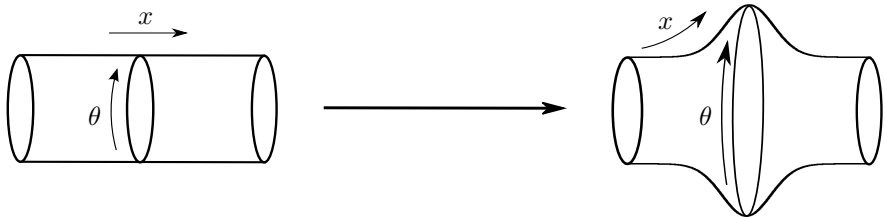}
		\put (14.5,24) {$x$}
		\put (71.5,23.5) {$x$}
		\put (9,13) {$\theta$}
		\put (76,13) {$\theta$}
	\end{overpic}
	\caption{Homeomorphism from a cylinder to a tube of varying radius}
	\label{homeo}
\end{figure}

This clearly contains Cartesian product manifolds as a subclass. Other than these, perhaps the simplest example is found by homeomorphically deforming a two-dimensional cylinder to get a tube of varying radius, as shown in Fig. \ref{homeo}.

Incidentally, as the metric for the cylinder, in the polar coordinates shown in Fig. \ref{homeo}, has the form
\begin{equation} \label{cylinder}
	\mathrm{g}_{IJ} = \left( \begin{array}{cc}
		1 & 0 \\
		0 & r^2
	\end{array} \right)
\end{equation}
we might na\"{i}vely assume that the metric for the tube of varying radius is 
\begin{equation}
	\mathrm{g}_{IJ} = \left( \begin{array}{cc}
		1 & 0 \\
		0 & r^2 (x)
	\end{array} \right)
\end{equation}
In fact, this is not the case. Varying the radius changes the distance traversed when moving in the $x$ direction too. By embedding the tube in three-dimensions, we can find the metric, and it turns out to be
\begin{equation}
	\mathrm{g}_{IJ} = \left( \begin{array}{cc}
		1 + (\frac{\partial r}{\partial x})^2 & 0 \\
		0 & r^2 (x)
	\end{array} \right)
\end{equation}
which can still be seen to reduce to (\ref{cylinder}) when $r$ is constant.

Note that because we have taken $r$ to vary only with $x$ and not with $\theta$, the metric satisfies the condition 
\begin{equation} \label{cyl-cond}
	\partial_Z \mathrm{g}_{\mu \nu} = 0
\end{equation}
This implies that the tube is a kind of manifold known as a `warped product'\cite{BO'N}.

(\ref{cyl-cond}) is the $N$-dimensional generalisation of Kaluza's `cylinder condition' for the metric\cite{Kaluza}. We will look at a consequence of this condition for covariant derivatives in Section \ref{CDs}.

However, for the most general product manifolds,
\begin{equation}
	\partial_Z \mathrm{g}_{\mu \nu} \neq 0
\end{equation}
and
\begin{equation}
	\partial_\rho \mathrm{g}_{XY} \neq 0
\end{equation}
There is therefore nothing that constrains the Levi-Civita connection to decompose as it did for Cartesian products: the holonomy group may not be $SO(t_1,s_1) \otimes SO(t_2,s_2)$. The Riemann tensor may therefore have components which are absent for Cartesian products. We will look more closely at the implications and interpretation of these features in Section \ref{Gauge}.

Nonetheless, we still have particular features associated with the case where the factor spaces are Einstein manifolds and two-dimensional manifolds. We pointed out in Section \ref{Cartesian} that for a Cartesian product of such manifolds, the covariant Ricci tensor is block diagonal in $y$-coordinates. Its operator form was diagonal, with all eigenvalues corresponding to each factor space being equal. In exactly the same way, we can see from condition (\ref{Ricci-Ein}) that for a \emph{general} product of Einstein manifolds or two-dimensional manifolds, the (covariant) Ricci tensor is block diagonal:
\begin{equation}
	R_{IJ} = \left( \begin{array}{cc}
		\frac{R_1}{N_1} \mathrm{g}_{\mu \nu} (y^\rho, y^Z) & 0 \\
		0 & \frac{R_2}{N_2} \mathrm{g}_{XY} (y^\rho, y^Z)
	\end{array} \right)
\end{equation}
and its operator form is
\begin{equation}
	R_I^J = \left( \begin{array}{cc}
		\frac{R_1}{N_1} \delta_\mu^\nu & 0 \\
		0 & \frac{R_2}{N_2} \delta_X^Y
	\end{array} \right)
\end{equation}
- that is, it is diagonal, with each of the eigenvalues associated with a given factor space being equal. Again, this holds in any coordinate system which respects the factor manifolds.

\section{Subgroups of $J$ and their associated coset decompositions \label{Subgroups}}

Having studied the geometry of product manifolds, we now want to look at the transformation properties of tensor fields on them. This requires a closer study of the group theory of basis changes on their tangent spaces.

Whenever a group has a subgroup, the subgroup can be used to partition the group into cosets. For linear Lie groups, these cosets form a coset space. Any group element can then be expressed as a product of a subgroup element and a `coset space representative'. For the group $J$ of all basis changes, we have several possible subgroups we could use for such a construction. 

\subsection{Decomposition using pseudo-orthogonal groups in $N$ dimensions \label{Sec-jli}}

One of these is the pseudo-orthogonal group in all $N$ dimensions, $I$. The generators of $I$ form a subset of the generators of $J$. Any element $j_0$ of $J$ may then be decomposed uniquely into a representative $l_0$ of the coset space $J/I$ and an element of $I$:
\begin{equation} \label{jli}
	j_0 = l_0 \, i_0
\end{equation}
with $l_0$ generated by the generators of $J$ which are not generators of $I$. This decomposition underlies the analysis in \cite{TSS}. 

However, the most general $i_0$ may not be continuously connected to the identity, as $I$ has more than one connected component. The connected subgroup of $I$ is an invariant subgroup of $I$. This means that we can partition $I$ into a finite number of cosets of its connected subgroup $S$. For example, for a space with positive definite metric,
\begin{equation}
	I = \{S, S P \}
\end{equation}
where $S \cong SO(N)$ and $P$ is a single matrix with determinant $-1$, which we can take to be diagonal. Alternatively, for our familiar four-dimensional spacetime,
\begin{equation}
	I = \{S, S P, S T, S P T\}
\end{equation}
where $S \cong SO_0(1,3)$ and $P$ and $T$ are both matrices with determinant $-1$, which are conventionally given the forms
\begin{equation}
	P = \begin{pmatrix}
		1 & & & \\
		& -1 & & \\
		& & -1 & \\
		& & & -1
	\end{pmatrix}
\end{equation}
and
\begin{equation}
	T = \begin{pmatrix}
		-1 & & & \\
		& 1 & & \\
		& & 1 & \\
		& & & 1
	\end{pmatrix}
\end{equation}

In general, then, we can obtain a decomposition of $j_0$ into three factors: $l_0$, $s_0 \in S$ and an element of a discrete Abelian group (for example, the $D^4$ group $\{\mathbf{1}, P, T, PT \}$).

\subsection{Decomposition using general linear groups \label{decomp}}

For product manifolds, we have an alternative. Under changes of coordinate system which respect the factor manifolds, the action on the basis is an element of a subgroup $G = G_1 \otimes G_2 \otimes \ldots$ of $J$. This can be used to partition $J$ into a coset space $J/G$, so that any change of basis $j_0$ can be uniquely decomposed in the form
\begin{equation} \label{jLg}
	j_0 = L_0 \, g_0
\end{equation}
where $g_0 = g_1 \otimes g_2 \otimes \ldots \in G$. Given that $j_0$ relates the coordinate basis $\mathbf{e}_M$ to a frame basis $\hat{\mathbf{n}}_I$:
\begin{equation}
	\mathbf{e}_M = (j_0)_M{}^I \hat{\mathbf{n}}_I
\end{equation}
this decomposition can be used to define a new basis on the tangent space:
\begin{equation} \label{mgn}
	\mathbf{m}_K = (L_0^{-1})_K{}^M \mathbf{e}_M = (g_0)_K{}^I \; \hat{\mathbf{n}}_I
\end{equation}
If $g_0$ is an element of a direct product of two general linear groups $G_1$ and $G_2$, then we may write 
\begin{equation}
	\mathbf{m}_\mu = (g_1)_\mu{}^\nu \hat{\mathbf{n}}_\nu \, , \; 
	\mathbf{m}_X = (g_2)_X{}^Y \hat{\mathbf{n}}_Y
\end{equation}
where $g_1 \in G_1$ and $g_2 \in G_2$. Note that $g_1$ and $g_2$ are not necessarily pseudo-orthogonal, so the new basis is not in general a frame basis. Instead, $g_1$ and $g_2$ map the chosen frame basis $\hat{\mathbf{n}}_I$ to a coordinate basis $\mathbf{m}_\mu$ on the tangent spaces $T_\Omega \mathcal{M}_1$ and a coordinate basis $\mathbf{m}_X$ on $T_\Omega \mathcal{M}_2$. (In a Kaluza-Klein theory, $\mathcal{M}_1$ would be our familiar four-dimensional spacetime, while $\mathcal{M}_2$ would be a compact space, which itself may be a product space.) $G_1$ and $G_2$ represent the sets of all such maps. This can be seen by taking the inner products of these new bases: this gives us a metric with the form (\ref{block-met}). $L_0$ then carries out a further transformation to a coordinate basis $\mathbf{e}_M$ which does not respect the factor spaces.

We thus find a condition for the neighbourhood to coincide with a product space: 

\noindent
\emph{if $L_0$ can be consistently defined across a coordinate neighbourhood $\Omega$, then
	\begin{itemize}
		\item $\mathbf{m}_K$ is a basis for coordinates $y^\rho,y^Z$ across $\Omega$;
		\item the manifold coincides with a product manifold across $\Omega$;
		\item the metric takes the form (\ref{block-met}) in these coordinates.
\end{itemize} }

\begin{figure} [ht]
	\centering
	\begin{overpic}[width=1\linewidth]{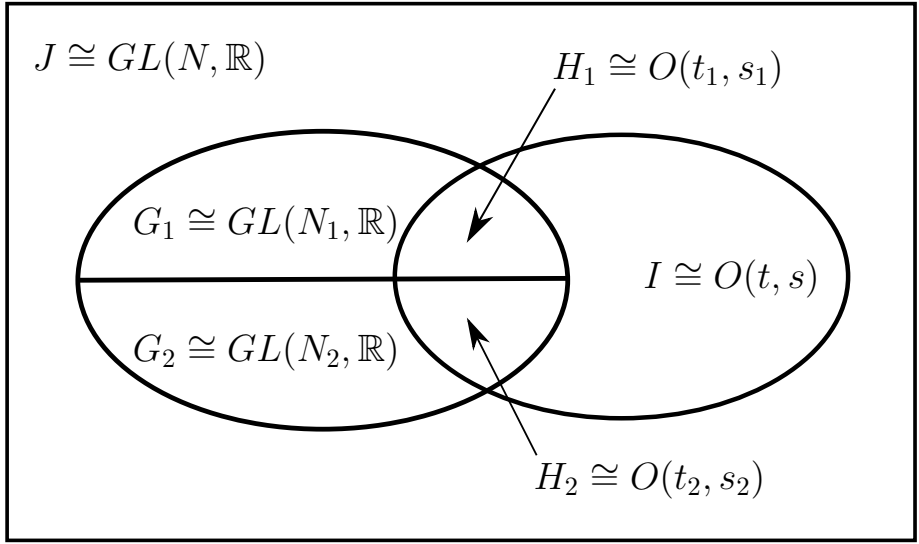}
		\put (3,52) {\Large$J \cong GL(N,\mathbb{R})$}
		\put (60,51) {\Large$H_1 \cong O(t_1, s_1)$}
		\put (14,34) {\Large$G_1 \cong GL(N_1,\mathbb{R})$}
		\put (70,28) {\Large$I \cong O(t,s)$}
		\put (14,20) {\Large$G_2 \cong GL(N_2,\mathbb{R})$}
		\put (58,6) {\Large$H_2 \cong O(t_2, s_2)$}
	\end{overpic}
	\caption{The relationships between tangent space groups on product manifolds}
	\label{groups}
\end{figure}

$G_1$ and $G_2$ have orthogonal or pseudo-orthogonal subgroups $H_1 \cong O(t_1,s_1)$ and $H_2 \cong O(t_2,s_2)$, as represented in Fig. \ref{groups}. (As always, this is readily extended to a product of more than two general linear groups.)

We will look into the physical relevance of these subgroups in Section \ref{Gauge}.

\subsection{Decomposition of tensors \label{tensor-decomp}}

On a neighbourhood which coincides with a product manifold - that is, one for which $L_0$ is consistently defined - we are able to apply the techniques of non-linear realisations\cite{CWZ}. The action of $j \in J$ on a coset $L_0 G$ maps it to another coset $L' G$, so
\begin{equation} \label{jLLg}
	j L_0 = L' g
\end{equation}
for some $g \in G$. This can be used to decompose all the higher-dimensional tensors on the space into tensors of the two factor spaces. For example, the coordinates of a vector in $y$-coordinates are
\begin{equation} \label{V-in-y}
	V_{(y)}^K = V_{(u)}^M (L_0)_M{}^K
\end{equation}
It is then easy to show that under the action of the full $J$, $V$ in $y$-coordinates transforms as a representation of $G$ only\cite{CWZ}:
\begin{equation}
	j: V_{(y)}^K \mapsto V_{(y')}^K = V_{(y)}^L (g^{-1})_L{}^K
\end{equation}
This breaks into multiplets of the factor groups; for example with two factor groups:
\begin{eqnarray}
	j: V_{(y)}^\mu &\mapsto& V_{(y')}^\mu = V_{(y)}^\nu (g^{-1})_\nu{}^\mu \label{V-sp} \\
	j: V_{(y)}^X &\mapsto& V_{(y')}^X = V_{(y)}^Y (g^{-1})_Y{}^X \label{V-int}
\end{eqnarray}
Higher rank tensors will break into blocks. For example, with two groups a covariant rank-two tensor field $Y_{IJ}$ will break into $Y_{\mu \nu }$, $Y_{\mu X }$, $Y_{X \nu }$ and $Y_{XY}$, transforming as
\begin{eqnarray}
	j: Y^{(y)}_{\mu \nu} &\mapsto& Y^{(y')}_{\mu \nu} = g_\mu{}^\rho g_\nu{}^\lambda Y^{(y)}_{\rho \lambda} \\
	j: Y^{(y)}_{\mu X} &\mapsto& Y^{(y')}_{\mu X} = g_\mu{}^\rho g_X{}^Y Y^{(y)}_{\rho Y} \\
	j: Y^{(y)}_{X \mu} &\mapsto& Y^{(y')}_{X \mu} = g_X{}^Y g_\mu{}^\rho Y^{(y)}_{Y \rho} \\
	j: Y^{(y)}_{W X} &\mapsto& Y^{(y')}_{W X} = g_W{}^Y g_X{}^Z Y^{(y)}_{Y Z}
\end{eqnarray}
In the Kaluza-Klein picture, these would be an uncharged rank-two tensor field, charged Lorentz vectors and a Lorentz scalar which transforms as the rank-two tensor representation of the internal symmetry.

\section{$O(N)$ gauge fields and the Riemann tensor \label{Gauge}}

We have just seen how the structure group of the tangent bundle is non-linearly realised and what this implies for the transformation of tensors. We can now look at what this implies for the Levi-Civita connection which appears in covariant derivatives, and its field strength, the Riemann tensor. We will see the appearance of $SO(s_2)$ gauge fields which couple to vector fields through the covariant derivatives. In the cases of two and three additional dimensions, we will see that these can be interpreted as gauge fields of $U(1)$ and $SU(2)$, acting on the doublet and triplet representations of these groups respectively.

\subsection{The Levi-Civita connection and its holonomy and intrinsic curvature}

\subsubsection{Minkowski spacetime and Cartesian product spaces}

In Minkowski spacetime, a system of coordinates may be used which has an orthonormal basis everywhere. The metric in these coordinates is the Minkowski metric and consequently the Levi-Civita connection is zero. It is then possible to induce inertial or `fictitious' forces by a change of coordinate system\cite{TSS}. In these new coordinates, the Levi-Civita connection will be non-zero, but its field strength, the Riemann tensor, remains zero, as the spacetime still has no intrinsic curvature. We can think of the Levi-Civita connection as like a gauge field here which can be `gauged away' by a change of coordinates (a return to the Minkowski coordinates) - it is `pure gauge'.

Naturally, this also applies to the Minkowski spacetime in the classical vacuum of our Kaluza-Klein theory. We have already noted that in a Cartesian product space, in $y$-coordinates, the Levi-Civita connection will decompose in the same way as the metric. This will obviously also be the case for a coordinate neighbourhood which is isometric to a Cartesian product. The coordinates $y^\mu$ on the four-dimensional spacetime may be chosen to be Minkowski coordinates. The Levi-Civita connection components $\mathring{\Gamma}_{\mu \nu}{}^\rho$ in these coordinates will all be zero. In any other set of coordinates $y'^\mu$ on this spacetime, this part of the Levi-Civita connection will be non-zero but still pure gauge. 

Under a change of coordinates which does not respect the factor spaces - to our $u$-coordinates, with basis $\mathbf{e}_I$ - the metric will have off-block-diagonal components. Indeed, this is the case for the Cartesian product of any subspaces. An example can be seen in the Klein metric (\ref{K-met}). If these vary with the coordinates, there will be new components to the Levi-Civita connection, such as $\mathring{\Gamma}_{\mu Y}{}^Z$. But such a change of coordinates does not change the intrinsic curvature, so these new components are again pure gauge.

This can be understood in terms of geodesics. By considering parallel transport round closed loops on a coordinate neighbourhood of the Cartesian product space, we can see that the holonomy group is a direct product of pseudo-orthogonal groups. More precisely, it is the group $H = G \cap I$ of basis changes which both preserve the frame metric and respect the factor spaces. (See Fig. \ref{groups} above.) Each of its factor groups $H_1 \subset G_1, H_2 \subset G_2, \ldots$ is the holonomy group for one of the factor spaces. Then by the de Rham decomposition theorem, (or rather, Wu's extension of it to manifolds with indefinite signature\cite{Wu}), geodesics are confined to these factor spaces. 

\subsubsection{More general product spaces}

The more general product spaces are qualitatively different. For these, the Levi-Civita connection has the additional components even in the $y$-coordinates. In general, a Levi-Civita connection has the following well-known form in terms of the metric:
\begin{equation}
	\mathring{\Gamma}_{IJ}{}^K = \mathring{\Gamma}_{JI}{}^K = \frac{1}{2} \mathrm{g}^{KL} \left(\partial_L \mathrm{g}_{IJ} - \partial_I \mathrm{g}_{LJ} - \partial_J \mathrm{g}_{LI} \right)
\end{equation}
The block diagonal form (\ref{block-met}) of the metric means that the additional components are
\begin{eqnarray}
	\mathring{\Gamma}_{\mu X}{}^Y &=& \mathring{\Gamma}_{X \mu}{}^Y = - \frac{1}{2} \mathrm{g}^{YZ} \partial_\mu \mathrm{g}_{ZX} \\
	\mathring{\Gamma}_{X Y}{}^\nu &=& \frac{1}{2} \mathrm{g}^{\nu \rho} \partial_\rho \mathrm{g}_{XY} \\
	\mathring{\Gamma}_{\mu X}{}^\nu &=& \mathring{\Gamma}_{X \mu}{}^\nu = - \frac{1}{2} \mathrm{g}^{\nu \rho} \partial_X \mathrm{g}_{\rho \mu} \label{compt3} \\
	\mathring{\Gamma}_{\lambda \mu}{}^X &=& \frac{1}{2} \mathrm{g}^{XZ} \partial_Z \mathrm{g}_{\lambda \mu} \label{compt4}
\end{eqnarray}
These represent additional intrinsic curvature. This can be seen in the toy example in Fig. \ref{homeo} above: the cylinder on the left-hand side has no intrinsic curvature, but the tube of varying radius on the right does. For these product spaces, the holonomy group is not $G \cap I$. Physically, this means that the geodesics are not confined to the factor spaces. Changing to the $u$-coordinates does not affect this: these are real components of curvature and the new components of the Levi-Civita connection cannot be gauged away.

\subsection{Covariant derivatives of vectors \label{CDs}}

The additional components of the Levi-Civita connection for product manifolds make an appearance in the covariant derivatives of vectors and tensors. In particular, for the more general product manifolds (in which they are not pure gauge), the covariant derivatives of $V^\mu$ and $V^Y$ with respect to the $y^\mu$ coordinates take the forms
\begin{eqnarray}
	\mathring{D}_\mu V^\nu &=& \partial_\mu V^\nu + V^\rho \, \mathring{\Gamma}_{\mu \rho}{}^\nu + V^X \, \mathring{\Gamma}_{\mu X}{}^\nu \\
	\mathring{D}_\mu V^X &=& \partial_\mu V^X + V^Y \, \mathring{\Gamma}_{\mu Y}{}^X + V^\nu \, \mathring{\Gamma}_{\mu \nu}{}^X
\end{eqnarray}
Note that when the condition (\ref{cyl-cond}) is satisfied, the Levi-Civita components (\ref{compt3}) and (\ref{compt4}) reduce to zero, so these covariant derivatives simplify to
\begin{eqnarray}
	\mathring{D}_\mu V^\nu &=& \partial_\mu V^\nu + V^\rho \, \mathring{\Gamma}_{\mu \rho}{}^\nu \\
	\mathring{D}_\mu V^X &=& \partial_\mu V^X + V^Y \, \mathring{\Gamma}_{\mu Y}{}^X \label{internal-d}
\end{eqnarray}
The transformation properties of the connection components in these expressions are of particular interest to us. Under an arbitrary change of coordinates, they transform according to
\begin{equation} \label{j-on-conn}
	j(u): \mathring{\Gamma}^{(u)}_{\lambda \mu}{}^\nu \mapsto \mathring{\Gamma}^{(u')}_{\lambda \mu}{}^\nu =  j_\lambda{}^\kappa \left( j \mathring{\Gamma}^{(u)}_\kappa j^{-1} \right)_\mu{}^\nu - j_\lambda{}^\kappa \left( j \partial_\kappa j^{-1} \right)_\mu{}^\nu
\end{equation}
where
\begin{equation}
	j_K{}^L(u)  = \frac{\partial u^L}{\partial u'^K}
\end{equation}
It is clear that $\mathring{\Gamma}_{\mu \rho}{}^\nu$ transforms in the normal way under changes of coordinates on the factor space $\mathcal{M}_1$ with $y^\mu$ coordinates and is invariant under changes of coordinates on the additional factor space $\mathcal{M}_2$ . More interestingly, under an element of $G_1 \otimes G_2$, $\mathring{\Gamma}_{\mu Y}{}^X$ transforms in the following way:
\begin{equation}
	\mathring{\Gamma}^{(u)}_{\mu X}{}^Y \mapsto \mathring{\Gamma}^{(u')}_{\mu X}{}^Y 
	= (g_1)_\mu{}^\nu \left(g_2 \, \mathring{\Gamma}^{(u)}_\nu g_2^{-1} - g_2 \, \partial_\nu g_2^{-1} \right)_X{}^Y
\end{equation}
- we see that under changes of coordinates on $\mathcal{M}_1$ it transforms as a covariant vector, while under changes of coordinates on $\mathcal{M}_2$ it transforms as a gauge potential. 

For a Kaluza-Klein theory, this looks promising: we have the appearance of a gauge field associated with a symmetry of the compact space. However, this is a gauge field of $G_2$ - a general linear group, not a compact group.

Now every connection with coordinate indices has an associated Lorentz or spin connection\cite{Pereira}. These were examined in \cite{TSS}, where it was shown that under a local change of frame, they transform as gauge fields of the group $I$ under which the frame metric is invariant:
\begin{equation} \label{Lgauge}
	i(u): \mathring{\omega}_{\mu \xi}{}^\rho \mapsto \mathring{\omega}'_{\mu \xi}{}^\rho =  \left( i \mathring{\omega}_\mu i^{-1} \right)_\xi{}^\rho + \left( i \partial_\mu i^{-1} \right)_\xi{}^\rho
\end{equation}
Thus $\mathring{\Gamma}^{(u)}_{\mu X}{}^Y$ has an associated Lorentz connection:
\begin{equation} \label{spin-cxn}
	\mathring{\omega}_{\mu X}{}^Y = (l_2^{-1})_X{}^W \Gamma^{(u)}_{\mu W}{}^V (l_2)_V{}^Y + (l_2^{-1})_X{}^W \partial_\mu (l_2)_W{}^Y \, .
\end{equation}
where $l_2$ is a representative of the coset space $G_2 / H_2$ whose generators do not contain any of the generators of $H_2$. (Its Greek index is therefore in a coordinate basis and the Roman indices are in a frame basis.) This transforms as a gauge field of $H_2 \cong O(s_2)$:
\begin{eqnarray}
	h_2 (u): \mathring{\omega}_{\mu X}{}^Y &\mapsto& \mathring{\omega}'_{\mu X}{}^Y =  \left( h_2 \mathring{\omega}_\mu h_2^{-1} \right)_X{}^Y + \left( h_2 \partial_\mu h_2^{-1} \right)_X{}^Y \\
	g_1 (u): \mathring{\omega}_{\mu X}{}^Y &\mapsto& \mathring{\omega}'_{\mu X}{}^Y = (g_1)_\mu{}^\nu \mathring{\omega}_{\nu X}{}^Y
\end{eqnarray}

This is very similar to the situation in the papers on spontaneous compactification by Volkov, Sorokin and Tkach. In these papers, the field equations have a solution in which the gauge fields are equal to the Lorentz connections associated with the Levi-Civita connection on the internal space\cite{VT80,VST83}. There is a slight difference: they identify the gauge group as the holonomy group of the Levi-Civita connection on the compact subspace. In our case, $H_2$ is only the holonomy group in the limit of a Cartesian product - that is, in the limit that the gauge field vanishes (or at least become pure gauge). But it should be noted that our model is very different from Klein's theory - it may be seen from (\ref{A-Klein}) that $A_\mu$ in Klein's theory is actually a set of four Jacobian matrix elements.

\medskip

The field strength of the gauge field (with respect to the four-dimensional spacetime coordinates) is
\begin{equation} \label{F-spin}
	\mathring{F}_{\mu \nu X}{}^Y = \partial_\mu \mathring{\omega}_{\nu X}{}^Y - \partial_\nu \mathring{\omega}_{\mu X}{}^Y + \mathring{\omega}_{\mu X}{}^Z \mathring{\omega}_{\nu Z}{}^Y - \mathring{\omega}_{\nu X}{}^Z \mathring{\omega}_{\mu Z}{}^Y
\end{equation}
Now Utiyama\cite{Utiyama} showed that such a field strength is related to the Riemann tensor. By substituting (\ref{spin-cxn}) into (\ref{F-spin}), a lengthy but straightforward calculation (easiest when the indices relating to the compact space are suppressed) shows that this field strength is indeed related to some of the components of the Riemann tensor:
\begin{equation} \label{RtoF}
	\mathring{F}_{\mu \nu X}{}^Y = (l_2^{-1})_X{}^Z R^V{}_{Z \mu \nu} \, (l_2)_V{}^Y 
\end{equation}

\medskip

Now, the gauge groups of the standard model are special unitary groups. It may therefore be wondered if our orthogonal gauge group has any physical relevance. But it should be remembered, as was explained in Section \ref{Intro}, that just as orthogonal transformations preserve the orthonormality of the frame basis, so unitary transformations preserve orthonormality on a complex vector space - the space inhabited by a spinor multiplet. 

Further work is needed to understand the relation between the unitary transformations and the orthogonal ones in detail in the general case, and following on from this, how to couple fermions directly to this model. However, for the two simplest cases, there are homomorphisms we can make use of to understand how particles described by vector fields couple to the model.

\subsection{Example: $s_2 = 2$}

Orthonormality on a vector space with one complex dimension is preserved under the group $U(1)$. This has a doublet representation which is the vector representation of $SO(2)$. 

In general, we write the generators of $SO(s_2)$ as 
\begin{equation} \label{gen'r}
	(M_{WZ})_X{}^Y = \mathrm{i} (\delta_W^Y \delta_{ZX} - \delta_{WX} \delta_Z^Y)
\end{equation}
so that in this case,
\begin{equation}
	M_{55} = M_{66} = 0; \hspace{10mm} M_{56}= - M_{65} 
	= \left( \begin{array}{cc}
		0 & -\mathrm{i} \\
		\mathrm{i} & 0
	\end{array} \right)
\end{equation}

Now $\mathring{\omega}_{\mu X}{}^Y$ is an element of $so(N)$. We will write it
\begin{equation}
	\mathring{\omega}_{\mu X}{}^Y = (\theta_\mu^{WZ} M_{WZ})_X{}^Y
\end{equation}
where $\theta_\mu^{WZ}$ is antisymmetric on its upper indices. Then substituting in (\ref{gen'r}) we find
\begin{equation}
	\mathring{\omega}_{\mu X}{}^Y = -2 \mathrm{i} \theta_{\mu X}{}^Y
\end{equation}
so by using $l_2$ to put (\ref{internal-d}) into a frame basis, we have
\begin{equation}
	\mathring{D}_\mu V^Y = \partial_\mu V^Y -2 \mathrm{i} V^X \, \theta_{\mu X}{}^Y
\end{equation}
Then using $\delta^{XY}$ and $\delta_{XY}$, we find that $\mathring{D}_\mu V^5$ and $\mathring{D}_\mu V^6$ are coupled:
\begin{eqnarray}
	\mathring{D}_\mu V^5 &=& \partial_\mu V^5 + 2 \mathrm{i} V^6 \, \theta_{\mu 5}{}^6 \\
	\mathring{D}_\mu V^6 &=& \partial_\mu V^6 - 2 \mathrm{i} V^5 \, \theta_{\mu 5}{}^6
\end{eqnarray}
To uncouple them, define 
\begin{equation}
	\mathrm{V} = V^5 + \mathrm{i} V^6
\end{equation}
Then
\begin{equation}
	\mathring{D}_\mu \mathrm{V} = \partial_\mu \mathrm{V} + 2 \, \theta_{\mu 5}{}^6 \, \mathrm{V}
\end{equation}
while
\begin{equation}
	\mathring{D}_\mu \mathrm{V}^* = \partial_\mu \mathrm{V}^* - 2 \, \theta_{\mu 5}{}^6 \, \mathrm{V}^*
\end{equation}
If we take
\begin{equation}
	\mathrm{i} e A_\mu = 2 \theta_{\mu 5}{}^6
\end{equation}
then these take exactly the right form for the coupling of a complex scalar field to the electromagnetic field\cite{Ryder}.

\subsection{Example: $s_2 = 3$ \label{s_2=3}}

Orthonormality on a vector space with two complex dimensions is preserved under the group $U(2) \supset SU(2)$. $SU(2)$ has a triplet representation which is the vector representation of $SO(3)$.

In this case, $\mathring{\omega}_{\mu X}{}^Y$ is an element of $so(3)$. Using the antisymmetry of the generators and their associated parameters, we find
\begin{equation}
	\mathring{\omega}_{\mu X}{}^Y = 2 \, \theta_\mu{}^{56} (M_{56})_X{}^Y + 2 \, \theta_\mu{}^{67} (M_{67})_X{}^Y + 2 \, \theta_\mu{}^{75} (M_{75})_X{}^Y
\end{equation}
We may write this
\begin{equation}
	\mathring{\omega}_{\mu X}{}^Y = 2 \, \theta_\mu{}^i (M_i)_X{}^Y 
\end{equation}
where $i = 1,2,3$ and
\begin{eqnarray}
	M_{1} &=& M_{67} = \left( \begin{array}{ccc}
		0 & 0 & 0 \\
		0 & 0 & -\mathrm{i} \\
		0 & \mathrm{i} & 0
	\end{array} \right) \\
	M_{2} &=& M_{75} = \left( \begin{array}{ccc}
		0 & 0 & \mathrm{i} \\
		0 & 0 & 0 \\
		-\mathrm{i} & 0 & 0
	\end{array} \right) \\
	M_{3} &=& M_{56} = \left( \begin{array}{ccc}
		0 & -\mathrm{i} & 0 \\
		\mathrm{i} & 0 & 0 \\
		0 & 0 & 0
	\end{array} \right)
\end{eqnarray}
and
\begin{equation}
	\theta_\mu{}^1 = \theta_\mu{}^{67}; \hspace{10mm} 
	\theta_\mu{}^2 = \theta_\mu{}^{75}; \hspace{10mm}
	\theta_\mu{}^3 = \theta_\mu{}^{56}
\end{equation}
so by using $l_2$ to put (\ref{internal-d}) into a frame basis, we have
\begin{equation}
	\mathring{D}_\mu V^Y = \partial_\mu V^Y + 2 \, V^X \theta_\mu{}^i (M_i)_X{}^Y 
\end{equation}
Then if we take
\begin{equation}
	\mathrm{i} g A_\mu{}^i = 2 \theta_\mu{}^i
\end{equation}
then using the antisymmetry of the generators this becomes
\begin{equation}
	\mathring{D}_\mu V^Y = \partial_\mu V^Y - \mathrm{i} g A_\mu{}^i (M_i)^Y{}_X V^X  
\end{equation}
- exactly the right form for the coupling of an $SU(2)$ gauge field to a multiplet of its vector representation\cite{Ryder}.

\subsection{Decomposition of the Riemann tensor}

We have seen how a product manifold whose factor spaces are two-dimensional manifolds and Einstein manifolds has a block diagonal Ricci tensor. This is true regardless of whether it is a Cartesian product or a more general product. It might initially seem odd that in homeomorphically deforming such a Cartesian product to a more general product, we are introducing new intrinsic curvatures, yet the Ricci tensor remains block diagonal. However, it must be remembered that the Riemann tensor has more degrees of freedom than the Ricci tensor. As the Riemann tensor is the field strength of the Levi-Civita connection, and it describes the intrinsic curvature of a manifold, it makes sense to look at its degrees of freedom for product spaces. We do that in this section.

On a product manifold, the components of the Riemann tensor can be separated into four classes, two of which may be further subdivided, as follows.
\begin{enumerate}
	\item Those with all indices of the same type: a) $R_{\mu \nu \rho \lambda}$, b) $R_{W X Y Z}$.
	\item Those with three indices of one type and one of the other: a) $R_{\mu \nu \rho X}$, b) $R_{\mu X Y Z}$. (All other components of this type are related to these two by symmetries of the tensor.)
	\item Those with the first pair of indices of one type and the other two of the other: $R_{\mu \nu X Y} = R_{X Y \mu \nu}$.
	\item Those with each pair of indices containing a mix of the two types:
	
	$R_{\mu X \nu Y} = - R_{X \mu \nu Y} = - R_{\mu X Y \nu} = R_{X \mu Y \nu}$.
\end{enumerate}

For a Cartesian product manifold in $y$-coordinates, only class 1 is non-zero. The contraction to get the Ricci tensor is on the first and third indices, so these components contribute to the Ricci tensor:
\begin{equation}
	R_{\nu \lambda} = \mathrm{g}^{\mu \rho} R_{\mu \nu \rho \lambda} \, , \; 
	R_{XZ} = \mathrm{g}^{WY} R_{WXYZ}
\end{equation}

For a more general product manifold in $y$-coordinates, the other classes may be non-zero.

As the metric is block diagonal in $y$-coordinates, Class 2 components only contribute to $R_{\mu X} = R_{X \mu}$ on contraction with it. 

Class 3 components cannot contribute to the Ricci tensor in $y$-coordinates - crucially, these are the components that are related to the field strength of the gauge fields through (\ref{RtoF}). \emph{Thus promoting the gauge fields to real, dynamical fields does not directly change the Ricci tensor in $y$-coordinates}. (Though it may do through field equations.)

Class 4 components only contribute to the Ricci tensors for the factor spaces (not to $R_{\mu X} = R_{X \mu}$) in $y$-coordinates:
\begin{eqnarray}
	R_{\nu \lambda} &=& \mathrm{g}^{\mu \rho} R_{\mu \nu \rho \lambda} + \mathrm{g}^{XY} R_{X \nu Y \lambda} \\
	R_{XZ} &=& \mathrm{g}^{WY} R_{WXYZ} + \mathrm{g}^{\mu \nu} R_{\mu X \nu Z}
\end{eqnarray}

In a more general system of coordinates, Class 1 components still only contribute to the Ricci tensors for the factor spaces. However, with a non-block diagonal metric, Class 2 and Class 4 components can now contribute to $R_{\mu \nu}$, $R_{XY}$ and $R_{\mu X} = R_{X \mu}$. Also, Class 3 can contribute to the off-block-diagonal parts of the Ricci tensor, $R_{\mu X} = R_{X \mu}$.

\section*{Part II: Orbits of $GL(N,\mathbb{R})$ \label{Orbits}}

We have seen how, on a spacetime which locally takes a product form, gauge fields of internal symmetries naturally arise. When a generalisation of Kaluza's cylinder condition is satisfied, vector fields are correctly coupled to these gauge fields. The classical vacuum of such a theory is a Cartesian product of a four-dimensional Minkowski spacetime and an Einstein manifold or two-dimensional manifold (or set of these). 

In the limit that \emph{all} curvature vanishes, we get a flat $N$-dimensional spacetime. (This is the limit $R_1, R_2 \rightarrow 0$ of (\ref{Ric2}), or the corresponding limit for the extension of this to more than two factor spaces.) On such a spacetime, the full $J \cong GL(N,\mathbb{R})$ symmetry is linearly realised.

What we would like to understand better at this point is the relationship between the Kaluza-Klein spacetime and the totally flat spacetime. Is there a coordinate-independent way of specifying a spacetime with a particular gauge symmetry? If so, can this be used to construct a mechanism for breaking the $J$-symmetry to its $I$-symmetry subgroup?

To examine this, it is helpful first to understand how internal symmetries are broken or non-linearly realised. We recap this in the next section.

\section{Breaking and non-linearly realising symmetries - a brief review \label{NLRSSB}}

\subsection{Internal symmetries}

1960 saw the publication of a paper by Gell-Mann and L\'{e}vy\cite{G-ML} which included a model which has since become known as the non-linear sigma model. This is a particularly simple case of a non-linear realisation of an internal symmetry group. In the same year (indeed, in the same journal), Goldstone\cite{Goldstone} introduced the concept of spontaneous symmetry breaking (SSB) in field theory. However, these topics were studied in such different ways that it was not proved until after nearly a decade of research that non-linear realisations such as the non-linear sigma model are the low-energy limits of field theories in which internal symmetries are spontaneously broken.

\medskip

In the non-linear sigma model, the norm of a multiplet of scalar fields is constrained, which allows one of the fields to be eliminated from the Lagrangian. For example, in the three dimensional case, we may write the norm of the multiplet as
\begin{equation}
	\phi^i \phi_i = r^2
\end{equation}
If we fix $r$, then we get a sphere in field space. We can then eliminate $\phi^3$ from the Lagrangian by writing it as 
\begin{equation}
	\phi^3 = (r^2 - \phi^a \phi_a)^{\frac{1}{2}}
\end{equation}
(Gell-Mann and L\'{e}vy actually considered the four-dimensional case and used the negative square root.) We then find that rotations in field space about the axis associated with $\phi^3$ are linearly realised, but all other rotations are non-linearly realised.

Gell-Mann and L\'{e}vy's paper was succeeded by many others during the 1960s in which various groups, particularly chiral groups, were non-realised\cite{chiral}. It was not until 1969 that a general mathematical framework for such non-linear realisations was presented, in a pair of papers by Callan, Coleman, Wess and Zumino\cite{CWZ,CCWZ}. These showed how a coset decomposition of a linear Lie group $G$ could be used to find the most general form of a Lagrangian in which a subgroup $H$ was linearly represented but the rest of the symmetries were realised non-linearly.

\medskip

Goldstone's paper and a follow-up with Salam and Weinberg\cite{GSW} looked at potentials with degenerate minima constructed out of scalar fields. In such models, the Lagrangian is invariant under a continuous global symmetry group but this invariance is not (fully) shared by its vacuum states. 

This mechanism was adapted by Higgs for a gauge symmetry\cite{Higgs1, Higgs2, Higgs3}, but the non-Abelian case was addressed by Kibble\cite{Kibble}. This paper again used a coset decomposition of the invariance group of the Lagrangian. It pointed out that the vacuum manifold could be identified with the coset space. This led researchers to realise that non-linear realisations represented the low-energy effective theory where a global symmetry was spontaneously broken\cite{Honerkamp,SS1}.

For example, if we take the famous Mexican hat potential:
\begin{equation}
	V = a^2 (\phi^i \phi_i - r^2)^2
\end{equation}
this is minimised for exactly the spherical field space described above. 

More generally, the potential is a function of a set of Lorentz scalar fields, which transform under a linear representation of the invariance group of the Lagrangian density, usually denoted $G$. One point in this field space is selected as the physical vacuum. This point is stabilised by a subgroup $H \in G$ which is remains unbroken at low energies. The vacuum manifold is then diffeomorphic to the coset space $G/H$. 

Now, in the non-linear sigma model, the constraint used is to fix the norm of the field multiplet. This results in a spherical field space, $S^N \approx SO(N+1) / SO(N)$. However, in electroweak theory or in unification based on Grand Unified Theories (GUTs), both $G$ and $H$ are unitary groups. The representation of $G$ chosen for the Goldstone or Higgs multiplet depends on the intended symmetry breaking pattern\cite{Isham3}. 

In the particular case in which $G$ is a unitary group and the transforms as the antisymmetric tensor representation of $G$, each value of this multiplet may be associated with an element of the Lie algebra $\mathcal{G}$ of $G$. The action of $G$ on $\mathcal{G}$ is by conjugation and this action partitions the algebra into orbits of the same eigenvalues\cite{MR,O'R2}. Each element of a given orbit has the same stabiliser group up to conjugation, determined by the multiplicities of its eigenvalues. The orbits may then be grouped into strata according to their stabiliser groups. If we are interested in breaking $G$ to a subgroup $H$, we pick a diagonal element of $\mathcal{G}$ which has stabiliser group $H$ to represent the physical vacuum. This is invariant under $H$, while elements of $G/H$ map it to other vacuum states. Consequently, as stated above, the vacuum manifold is diffeomorphic to $G/H$.

An orbit may be specified by the eigenvalues of the elements of $\mathcal{G}$ it contains, or equivalently, by a set of invariants appearing in its characteristic equation. To get a chosen symmetry breaking pattern, we construct a potential for the Goldstone or Higgs multiplet which is minimised for the corresponding values of these invariants. (This is a generalisation of the Mexican hat potential being constructed to be minimised for a chosen value of the invariant $\phi^i \phi_i$.)

\subsection{Spacetime symmetries}

The analysis reported in this part of this paper is based on adapting the theory for internal symmetries to changes of basis on tangent spaces. However, during the writing of this paper, it has come to the author's attention that various researchers have already considered aspects of this down through the decades. Here, we give a brief account of the relevant papers that the author is currently aware of - no claim for completeness is made.

The history here goes back even further. Rainich\cite{Rainich} and Churchill\cite{Churchill} considered the algebraic properties of antisymmetric and symmetric rank-two tensors in 1925 and 1931 respectively. They did so in a way which was independent of coordinate system (and therefore invariant under changes of coordinates), by looking at eigenspaces of these tensors. As Churchill's paper focused on symmetric tensors, it is highly relevant to the study below. A key point in its analysis is that, as we shall see, not every such tensor is diagonalisable. In many cases, the matrices of components can only be reduced to Jordan normal form. These can be classified into Segre types and Churchill identified a 'canonical form' for each type.

This is a key difference from the situation described above for unitary symmetries, and underlies many of the later papers, which tend to carry out analysis which is valid for all Segre types. Analysis of Segre types and canonical forms has been extended to higher numbers of dimensions - see, for example, Rebou\c{c}as \emph{et al}\cite{RST}.

As we have seen, an index-aligned tensor can be seen as an operator acting on a vector, and an element of $GL(N,\mathbb{R})$ representing a change of coordinate basis acts on it by conjugation. The mathematics of how matrices used in such linear maps transform under similarity transformations was studied by Arnold\cite{Arnold}, who showed that these transformation partition the matrices into orbits. Each orbit, as we shall see, has an associated stabiliser group (centraliser). This allows the orbits to be collected into `z-classes' (these appear to be the `strata' referred to above); this has been examined for the case of the action under $GL(N,\mathbb{R})$ by Kulkarni\cite{Kulkarni}. Further references to this body of research can be found in Cirici\cite{Cirici}.

Another paper which is highly relevant is that by Hervik and Coley\cite{HC}. Like many papers on this subject, it focuses on curvature tensors. (It actually looks not just at the rank-two tensor operators, but at forming operators from curvature tensors of any rank.) It looks at how the diagonalisability of these operators relates to the extent to which the metric can be determined by their invariants.

\section{Subspaces of $gl(N,\mathbb{R})$ \label{Subspaces}}

The starting point for this study is to note that the components of any real rank-two tensor in operator form  at $A$ form a real $N \times N$ matrix and therefore an element of $\mathcal{J}_A \cong gl(N,\mathbb{R})$, the Lie algebra of $J_A$. (We will go back to looking at a single tangent space for now, and consider field-valued versions of the tensor and transformations in Section \ref{Constraints}.)

This Lie algebra splits naturally into two orthogonal subspaces, the $\mathcal{I}_A \cong o(t,s)$ subalgebra, and the subspace which may be exponentiated to obtain representatives of the coset space $\mathcal{J}_A / \mathcal{I}_A $. It is easy to show that matrices with the property (\ref{antisymm}) belong to the subalgebra $\mathcal{I}_A$, by showing that its exponential preserves the frame metric. Exponentiating $\bar{X} |_A$ gives us
\begin{equation} \label{jbarX}
	(j_{\bar{X}} |_A)_I{}^J \equiv \mathrm{exp} (\mathrm{i} \bar{X}_I{}^J |_A) \equiv \delta_I^J + \mathrm{i} \bar{X}_I{}^J |_A + \frac{1}{2} \bar{X}_I{}^K  |_A \bar{X}_K{}^J  |_A + \dots
\end{equation}
so that 
\begin{equation}
	j_{\bar{X}} |_A: \eta_{IJ} \mapsto \mathrm{g}_{IJ} |_A = (\delta_I^K + \mathrm{i} \bar{X}_I{}^K |_A) (\delta_J^L + \mathrm{i} \bar{X}_J{}^L |_A) \eta_{KL} |_A + \mathcal{O}^2 (\bar{X}_I{}^J |_A)
\end{equation}
This preserves $\eta_{IJ}$ if and only if
\begin{equation}
	\mathrm{i} \bar{X}_I{}^K |_A \eta_{KJ} |_A + \mathrm{i} \bar{X}_J{}^L |_A \eta_{IL} |_A = 0
\end{equation}
and hence
\begin{equation}
	\bar{X}_I{}^M  |_A= - \eta^{MJ} |_A \bar{X}_J{}^L |_A \eta_{IL} |_A = - \bar{X}^M{}_I
\end{equation}
as stated. Conversely, if $X |_A$ is an index-aligned tensor, in the frame basis it lies in the subspace of $\mathcal{J}_A$ orthogonal to $\mathcal{I}_A$. 

\section{The action of $GL(N,\mathbb{R})$ on index-aligned tensors - diagonalisable and non-diagonalisable orbits \label{diag-non}}

The action (\ref{j-action}) on the space of all real index-aligned tensors is therefore the action of a group on a subspace of its own Lie algebra. It is an action by conjugation, so it preserves eigenvalues - that is, $X_I^J |_A$ has the same eigenvalues in every coordinate system. This partitions the space into orbits, related by changes of basis on the tangent space. 

Each element of a given orbit has the same set of eigenvalues. For a space of positive definite metric, these eigenvalues are all real. For such a space, we can always diagonalise $X_I^J |_A$ using an element of $J_A$, as explained in the Appendix.

But for a spacetime of indefinite metric, the eigenvalues may in general be real or complex. Similarly, their corresponding eigenvectors may be either real or complex. For these spacetimes, whether $X_I^J |_A$ can be diagonalised depends on whether it lies in an orbit containing diagonal matrices. 

We now consider three classes of component matrices for real index-aligned tensors: i) those with some eigenvalues complex, ii) those with all eigenvalues real and distinct, and iii) those with all eigenvalues real and some of them repeated. For each class, we discuss whether the corresponding orbits contain diagonal matrices. For those that do, we will be able to adapt the analysis of Michel and Radicati\cite{MR}.

\subsection{Complex eigenvalues}

As $X |_A, j |_A$ and $j^{-1} |_A$ are all real matrices, $X' |_A$ must be real too. This means, as pointed out by Churchill\cite{Churchill}, that if $X |_A$ has any complex eigenvalues, it cannot be diagonalised by an element of $J_A$.

This is a significant difference from the action of a special unitary group on its Lie algebra\cite{MR}. The Lie algebra of $SU(N)$ is the set of all traceless Hermitian $N \times N$ matrices. A hermitian matrix may always be diagonalised by an appropriate unitary one and any matrix in $su(N)$ can be diagonalised by an appropriate element of $SU(N)$. 

Complexity of the eigenvalues puts a constraint on the eigenvectors. The eigenvalue equation for a real index-aligned tensor at a point $A$ on a spacetime with an arbitrary metric is
\begin{equation} \label{eval}
	X^I_J |_A t^J = \lambda t^I
\end{equation}
Premultiplying this by $t^*$, we have
\begin{equation}
	t^*_I X^I_J |_A t^J = \lambda t^*_I t^I \Rightarrow X^I_J |_A t^*_I t^J = \lambda t^*_I t^I \label{evec*}
\end{equation}
Now by taking the complex conjugate of (\ref{eval}), lowering the free index with the metric and remembering that $X^I_J |_A$ is real, we find
\begin{equation}
	X_{KJ} |_A t^{*J} = \lambda^* t^*_K \Rightarrow X_K^I |_A t^*_I = \lambda^* t^*_K
\end{equation}
Substituting this into (\ref{evec*}), we find
\begin{equation}
	\lambda^* t^*_J t^J = \lambda t^*_I t^I
\end{equation}
From this we see that the eigenvalue is real:
\begin{equation}
	\lambda = \lambda^*
\end{equation}
unless
\begin{equation} \label{c-null}
	t^*_I t^I = 0
\end{equation}
Thus complex eigenvalues can only occur when the corresponding eigenvector satisfies the `complex null' condition (\ref{c-null}).

\subsection{Real distinct eigenvalues}

Perhaps of even more relevance to us than the comparison with $su(N)$ is the comparison with the spectral theorem for symmetric matrices. This tells us that any symmetric matrix may be diagonalised using an orthogonal one. We can adapt the proofs of this to show that any index-aligned tensor with distinct real eigenvalues may be diagonalised using an element of $GL(N, \mathbb{R})$, as shown in the second part of the Appendix. A geometric interpretation of this is given in the third part of the Appendix.

\subsection{Real repeated eigenvalues}

We have already noted that two matrices of tensor components which are conjugate have the same eigenvalues. However, for spacetimes of indefinite metric, the converse is not true. Two matrices of tensor components may have the same eigenvalues but not be conjugate, if they have repeated real eigenvalues. This is another major difference from the orbit structure of $su(N)$. Some orbits contain diagonal matrices and others do not. Whether a matrix can be diagonalised cannot be determined by the eigenvalues alone - the eigenspaces need to be considered. We can illustrate this issue with the following matrices (we are not suggesting these are components of index-aligned tensors, merely providing an illustration of the issue)
\begin{equation}
	X |_A = \begin{pmatrix}
		12/5 & 2/5 & -4/5 \\
		0 & 2 & 0 \\
		-4/5 & -4/5 & 18/5 
	\end{pmatrix} ;
	\hspace{10mm}
	Y |_A = \begin{pmatrix}
		3 & 0 & 1 \\
		0 & 4 & 0 \\
		-1 & 0 & 1 
	\end{pmatrix}
\end{equation}
Both have eigenvalues $2,2,4$.

For $X |_A$, when $\lambda = 2$ is substituted into the eigenvalue equation, there are two non-trivial component equations, both of which reduce to
\begin{equation}
	t^1 + t^2 - 2t^3 = 0
\end{equation}
Thus the vector with components $(2, 0, 1)$ is an eigenvector, as is the vector with components $(0, 2, 1)$ - and indeed, any linear combination of these is an eigenvector. This double eigenvalue is therefore associated with a two-dimensional eigenspace: in the jargon, the eigenvalue has equal algebraic and geometric multiplicities. The eigenvalue $4$ also has an associated real eigenvector (which is independent of these - see the Appendix), so we can therefore choose three real independent eigenvectors for this $3 \times 3$ matrix, meaning that it can be diagonalised by a real, invertible matrix. 

For $Y |_A$, on the other hand, when $\lambda = 2$ is substituted into the eigenvalue equation, there are three non-trivial component equations; two of these reduce to
\begin{equation}
	t^1 - t^3 = 0
\end{equation}
while the other reduces to
\begin{equation}
	t^2 = 0
\end{equation}
Thus the vector with components $(1, 0, -1)$ is an eigenvector, as is any multiple of it. The double eigenvalue is therefore associated with a one-dimensional eigenspace: it has an algebraic multiplicity larger than its geometric multiplicity. Thus we can only find two independent real eigenvectors in total for $Y$, so it cannot be diagonalised.

As $X |_A$ is conjugate to $D |_A = \mathrm{diag} \, (2,4,2)$ but $Y |_A$ is not, these matrices are not in the same orbit, despite having the same eigenvalues.

\section{Cartan subspaces and stabilisers \label{Stab}}

We have seen that all orbits of matrices with real distinct eigenvalues contain at least one diagonal matrix, and some orbits of matrices with repeated real eigenvalues contain at least one diagonal matrix. 

The set of all diagonal matrices form an $N$-dimensional Abelian subalgebra of $gl(N, \mathbb{R})$. No other matrices commute with all of these, so it is a Cartan subalgebra. For any diagonal tensor $D |_A$ with distinct eigenvalues, its powers (including $(D |_A)^0 = \mathbf{1}$) up to $(D |_A)^{N-1}$ are linearly independent and are all themselves diagonal, so they span this diagonal Cartan subspace. The traces of these powers are also independent:
\begin{eqnarray}
	\mathrm{tr} (D |_A) &=& \sum_{I=0}^{N-1} \lambda_{(I)} \\
	\mathrm{tr} ((D |_A)^2) &=& \sum_{I=0}^{N-1} (\lambda_{(I)})^2 \\
	\ldots & & \\
	\mathrm{tr} ((D |_A)^{N-1}) &=& \sum_{I=0}^{N-1} (\lambda_{(I)})^{N-1}
\end{eqnarray}

For a diagonal tensor with repeated eigenvalues, these traces are not all independent.

Commutation is preserved under any inner automorphism:
\begin{equation}
	[j X j^{-1}, j Y j^{-1}] |_A = j [X,Y] j^{-1} |_A
\end{equation}
so if one acts on the whole diagonal Cartan subalgebra by conjugation, the result is a non-diagonal Cartan subalgebra. The powers and eigenvalues of a matrix are also preserved under conjugation, so if the diagonal tensor $D |_A$ has distinct eigenvalues, the Cartan subspace containing $X |_A = j |_A D |_A j^{-1} |_A$ is spanned by its powers. More on Cartan subspaces can be found in the Appendix.

\medskip

A tensor matrix with distinct eigenvalues is stabilised only by group elements generated by the Cartan subalgebra it is an element of. Matrices with repeated eigenvalues have a larger stabiliser group. 

For $gl(N,\mathbb{R})$, a diagonal tensor with distinct eigenvalues such as 
\begin{equation}
	\begin{pmatrix}
		1 & &  & & \\
		& 2 & & & \\
		&  & 3 & & \\
		&  & & \ddots & \\
		& & & & N
	\end{pmatrix}
\end{equation}
is stabilised by a group $GL(1,\mathbb{R})^N$. However, if two eigenvalues are the same, there is a $GL(2,\mathbb{R})$ factor in its stabiliser; if three eigenvalues are the same, there is a $GL(3,\mathbb{R})$ factor in its stabiliser, and so forth. The maximal stabiliser group is the whole of $J_A$. This is the stabiliser for a diagonal tensor with all eigenvalues the same, that is, a multiple of the Kronecker delta.

The eigenvalues are preserved under conjugation, so the image of a diagonal tensor (that is, the tensor in a different coordinate system), has the same stabiliser up to conjugation. Under this action, the Kronecker delta is invariant, so orbits for which the stabiliser is $J_A$ have only one element.

\section{The constraints on $X$ corresponding to product manifolds \label{Constraints}}

We are now in a position to answer the questions posed at the start of Part II. 

\emph{If \emph{any} real index-aligned tensor $X(u)$ is diagonalisable across a neighbourhood (open set) $\Omega$ of $\mathcal{M}$ and the multiplicities of its eigenvalues are the same across this neighbourhood, then its stabiliser is $\mathrm{Stab}_X = G$ across $\Omega$. This means that it can be diagonalised by a field-valued representative $L_0$ of the coset space $J/G$, which is consistently defined across $\Omega$.}

\emph{If it has at least two distinct eigenvalues, $G$ is a proper subgroup of $J$. Then, from the results of Section \ref{decomp}, the manifold coincides across $\Omega$ with some form of product manifold, on which we can define the $y$-coordinates with basis $\mathbf{m}_K$. The dimensionalities of the factor spaces are equal to the multiplicities of the eigenvalues of $X$ and $L_0{}^{-1}$ represents a change of basis from the basis for the $u$-coordinates to the basis for the $y$-coordinates.}

As pointed out in Section \ref{Structure}, such a decomposition into factor manifolds across a neighbourhood is far from unusual - it is those manifolds which \emph{cannot} be decomposed in this way which are the exceptions.

It is also worth remarking that this does not mean that all diagonalisable index-aligned tensors can be simultaneously diagonalised across $\Omega$. For example, $X$ may relate to a particular form of matter in the system. It may be that other index-aligned tensors relating to other forms of the matter in the system do not share this property. The existence of the $y$-coordinates means that they can be decomposed into tensors of $G$, but they need not be block diagonal, let alone completely diagonal.

These rather formal results can helpfully be illustrated for a Kaluza-Klein theory. Say that a real index-aligned tensor $X(u)$ can be diagonalised to $D = \mathrm{diag} (a, a, a, a, b, b, b)$ across a seven-dimensional neighbourhood $\Omega \in \mathcal{M}$, where the first of these eigenvalues corresponds to a timelike direction and the remainder to spacelike ones, and $a$ and $b$ are different real functions across $\Omega$. (This would be a tensor of Segre type $[(1,111)(111)]$ - see, for example, Rebou\c{c}as \emph{et al}\cite{RST}.) Then $G = Stab_X = G_1 \otimes G_2 \cong GL(4,\mathbb{R}) \otimes GL(3,\mathbb{R})$. $\mathcal{M}$ coincides across $\Omega$ with some form of product of $\mathcal{M}_1$ and $\mathcal{M}_2$, where $\mathcal{M}_1$ is a four-dimensional spacetime with Lorentzian signature and $\mathcal{M}_2$ is a three-dimensional space with positive definite signature. In such a situation, $X = D$ in any $y$-coordinates, while any other rank-two tensor $Y$ decomposes into four parts, $Y_{\mu \nu}$, $Y_{\mu X}$, $Y_{X \nu}$ and $Y_{XY}$, which transform as their indices suggest under $G_1$ and $G_2$. Tensors of other ranks similarly decompose into multiplets of $G_1$ and $G_2$. As we saw in Section \ref{s_2=3}, among the components of the Levi-Civita connection there are gauge fields of $SU(2) \rightarrowtail SO(3)$ (the symbol here denotes the homomorphism). The most general form of $X$, that is, in $u$-coordinates, can then be written
\begin{equation}
	X = L_0 D L_0{}^{-1}
\end{equation}
where $L_0$ is a representative of $J/G \cong GL(N,R) / (GL(4,\mathbb{R}) \otimes GL(3,\mathbb{R}))$. 

Similarly, on a nine-dimensional space, if $X$ diagonalises to 
\begin{equation} \label{4a3b2c}
	D = \mathrm{diag} (a, a, a, a, b, b, b, c, c)
\end{equation}
across $\Omega$, where $a,b,c$ are all different real functions, and again there is one timelike direction which is associated with the first eigenvalue, then $G = Stab_X = G_1 \otimes G_2 \otimes G_3 \cong GL(4,\mathbb{R}) \otimes GL(3,\mathbb{R}) \otimes GL(2,\mathbb{R})$. The model then contains gauge fields and tensors of $SU(2) \otimes U(1) \rightarrowtail SO(3) \otimes SO(2)$.

\medskip

The eigenvalues of $X$ are determined by its characteristic equation:
\begin{equation}
	(X - \lambda_0 \mathbf{1}) (X - \lambda_1 \mathbf{1}) \ldots = 0
\end{equation}
Multiplying out the factors gives a polynomial whose coefficients can be expressed in terms of the traces of the powers of $X$. A set of eigenvalues can therefore be specified by fixing the traces of these powers. These traces are always real and the traces of the even powers are always non-negative. For example, a matrix which diagonalises to (\ref{4a3b2c}) has the nine invariants
\begin{equation}
	\mathrm{tr} (X^I) = 4 a^I + 3 b^I + 2 c^I
\end{equation}
(where $I$ here represents powers from one to nine, rather than labelling coordinate directions).

As noted in the previous subsection, these will not all be independent, because of the repeated eigenvalues. It is easy to show, for example, that a matrix with the eigenvalues $(a, a, 0, 0, \ldots)$ satisfies
\begin{equation}
	\mathrm{tr} (X) \mathrm{tr} (X^2) = 2 \mathrm{tr} (X^3)
\end{equation}

Just as we had for the diagonal matrix in Section \ref{Stab}, if $X$ has all eigenvalues different, then $G \cong (GL(1,\mathbb{R}))^N$. At the other extreme, if all of its eigenvalues are equal, then its stabiliser is just $J$ and it may be written 
\begin{equation} \label{unbroken2}
	X_I^J = a \delta^J_I 
\end{equation}
in any coordinate system. This does not impose any product decomposition on the spacetime (although the diagonalisation of a different tensor across a neighbourhood might).

\medskip

Crucially, the traces of the powers of $X$ are coordinate-independent. We therefore have a coordinate-independent way of specifying a spacetime with a particular gauge symmetry\footnote{At least, up to changes of signature. More research is needed on this point. It is known, for example, that under an overall change of signature (one changing every timelike dimension into a spacelike one, and \emph{vice versa}), the sign of the Ricci scalar changes\cite{HC}. (See Section \ref{PRIs} for more on pure Ricci invariants.) However, it is currently unclear whether there are invariants of a particular tensor field which can be used to distinguish between, for example, an $SO(3)$ gauge symmetry on a four-dimensional spacetime with Lorentzian signature and an $SO(1,2)$ gauge symmetry on a four-dimensional spacetime of Lorentzian or positive definite signature. That is, the conditions under which the invariants may be different for Segre types $[(1,111)(111)]$, $[(1,111)(1,11)]$ and/or $[(1111)(1,11)]$.}. Using such algebraic invariants is essentially the method used in non-linear $\sigma$-models to define a field space on which internal symmetries are non-linearly realised (see, for example, Eichenherr\cite{Eichenherr}).

However, with an indefinite metric, using these invariants to specify the eigenvalues of an arbitrary real, index-aligned tensor is insufficient to ensure that spacetime coincides with a product manifold across a neighbourhood. It is also necessary to require that tensor to be diagonalisable across the neighbourhood. 

Now the modern view of non-linear sigma models is as the vacuum manifolds when internal symmetries are spontaneously broken. That is, a potential is constructed from (Lorentz) scalar fields, which is minimised when the invariants take the chosen values. The field states which satisfy these constraints then form a continuous vacuum manifold and the broken symmetries map between these states.

The problem with adapting this for a Kaluza-Klein theory is the following. With an indefinite metric, not all of the values of a real, index-aligned tensor with the required values of the invariants are related by coordinate changes. (That is, unless all the eigenvalues are distinct - in which case, the product manifold is a product of one-dimensional manifolds.) To ensure that we get a product manifold, and hence the desired gauge symmetries, we must also impose that $X$ is diagonalisable, and there is no obvious way to do this algebraically.

In the absence of this, we must impose at least the diagonalisability `by hand', that is as an ansatz for a solution of whatever field equations we are using.

\section{Constraining the Ricci tensor using the pure Ricci invariants \label{PRIs}}

Finally, it is worth reflecting on what happens when the constraints are applied to one of the tensors listed in Section \ref{Simplest}, as this has a bearing on the classical vacuum for our theory.

If we use the invariants to fix the eigenvalues for any of the three tensors $R_J^J$, $S_I^J$ or $G_I^J$ across a neighbourhood, or simply to fix their multiplicities, then this tensor is diagonal across the neighbourhood in $y$-coordinates. It is then easy to see from the definitions of $S_I^J$ and $G_I^J$ that we are effectively fixing the eigenvalues or their multiplicities for the other two as well. Indeed, if the Einstein field equations hold, then fixing the eigenvalues or their multiplicities for any of the four tensors $R_J^J, S_I^J, G_I^J, T_I^J$ fixes these for the other three, making all four of them diagonal across the chart.

In Section \ref{GPM}, we saw that for a product of Einstein manifolds and/or two-dimensional manifolds, the operator form of the Ricci tensor is diagonal in $y$-coordinates, with each of the eigenvalues associated with a given factor space being equal.

These new results mean that we can now prove the converse. Any manifold with the following properties:
\begin{itemize}
	\item the operator form of the Ricci tensor is diagonalisable across a neighbourhood
	\item it has the same multiplicities of eigenvalues at each point in that neighbourhood
\end{itemize}
\emph{necessarily} coincides with a product space of Einstein manifolds and/or two-dimensional manifolds across that neighbourhood\footnote{See Hervik and Coley\cite{HC} for more on the implications of diagonalisable curvature operators.}. The argument goes as follows. We know that if any real, index-aligned tensor has these properties, then the manifold coincides with a product space across the neighbourhood. This means we can use $y$-coordinates across it. In any such coordinate system, the operator form of the Ricci tensor remains diagonal, with each of the eigenvalues associated with a given factor space being equal, because it is invariant under $G = \mathrm{Stab}_R$. The covariant Ricci tensors for the factor spaces are therefore proportional to their metrics, so these factor spaces are Einstein manifolds or two-dimensional manifolds.

In more general coordinates, which do not respect the factor spaces, the Ricci tensor at each point across the coordinate neighbourhood will take a value in a $J$-orbit.  

As explained in Section \ref{Constraints}, the eigenvalues of a real index-aligned tensor may be specified by the traces of its powers. For the Ricci tensor, these are known as the pure Ricci invariants. The first such trace is the Ricci scalar, $R$. In four-dimensional spacetime, there are three more independent traces, which can be expressed in terms of $R$ and the traceless Ricci tensor $S_\mu^\nu$, and hence in terms of the first four Carminati-McLenaghan invariants, $r_1, r_2, r_3$\cite{CM2}:
\begin{eqnarray}
	R_\mu^\nu R_\nu^\mu &=& S_\mu^\nu S_\nu^\mu + \frac{R^2}{4} = 4 r_1 + \frac{R^2}{4} \\
	R_\mu^\nu R_\nu^\rho R_\rho^\mu &=& S_\mu^\nu S_\nu^\rho S_\rho^\mu  + \frac{3R}{4} S_\mu^\nu S_\nu^\mu + \frac{R^3}{16} = -8 r_2 + 3 R r_1 + \frac{R^3}{16} \\
	R_\mu^\nu R_\nu^\rho R_\rho^\sigma R_\sigma^\mu &=& S_\mu^\nu S_\nu^\rho S_\rho^\sigma S_\sigma^\mu + R S_\mu^\nu S_\nu^\rho S_\rho^\mu + \frac{3R^2}{8} S_\mu^\nu S_\nu^\mu + \frac{R^4}{64} \nonumber \\
	&=& 16 r_3 -8 r_2 + \frac{3 R^2}{2} r_1 + \frac{R^4}{64}
\end{eqnarray}

For an $N$-dimensional spacetime, we shall extend the notation of Narlikar and Karmarkar and Harvey\cite{Harvey}, to define
\begin{eqnarray}
	I_1 &=& R = R_I^I \\
	I_2 &=& R_I^J R_J^I \\
	I_3 &=& R_I^J R_J^K R_K^I \\
	\ldots
\end{eqnarray}
Specifying these across a chart $\Omega$ determines the eigenvalues of $R_I^J$ and hence its stabilisers. For example, if $N = 6$ and each of the invariants have the form 
\begin{equation}
	I_K = 2 a^K (u)
\end{equation}
(with $K$ labelling powers), then in $y$-coordinates
\begin{equation}
	R_I^J = \mathrm{diag} (0, 0, 0, 0, a(y), a(y)) = \mathrm{diag} (0, 0, 0, 0, R(y) / 2, R(y) / 2)
\end{equation}
Consequently, our manifold must coincide over $\Omega$ with a product space of a Ricci-flat four-dimensional spacetime and a Ricci-curved two-dimensional spacetime. In any coordinate system which respects the factor spaces (any $y$ coordinate system), we then have
\begin{equation}
	R_\mu^\nu = 0 \, , \; R_X^Y = \frac{R(u)}{2} \delta_X^Y
\end{equation}

\medskip

Note that in all of this analysis, the manifold is not required to be a product space globally. $\Omega$ should be selected according to the symmetries of a system. For example, in GR, the eigenvalues of $R_I^J$ depend on the properties of the matter at that point. The stabiliser of $R_I^J$ may therefore be different inside a matter distribution to that outside it.  (Also, there may be different types of matter present, with different symmetries - in this case, the \emph{total} energy-momentum density tensor determines the stabiliser of $R_I^J$.)

\section{Conclusions and discussion \label{Concs}}

We have investigated many aspects of product manifolds - both Cartesian products and more general products. While many of the results are valid for manifolds of any dimensionality and any signature, we have focused on their relevance to Kaluza-Klein theories.

We have seen that for a theory for which the field equations relate a real index-aligned tensor field describing the geometry to another describing the matter content, there are solutions which describe a product manifold. 

The decompactification limit of such a solution is a flat $N$-dimensional spacetime, with $N$ real coordinates. All of its tangent spaces are real vector spaces, all of its timelike dimensions are identical and all of its spacelike dimensions are identical. 

The solutions of most interest for Kaluza-Klein theories are those with a four-dimensional factor space of Lorentzian signature and a compact factor space of definite signature, where the four-dimensional spacetime metric is independent of the compact space coordinates, but the compact space metric varies with the spacetime coordinates. On such a space, we have seen that there are naturally $O(s_2)$ gauge fields present - these are the spin connections corresponding to certain components of the Levi-Civita connection. Tensor fields decompose into multiplets of this symmetry and the Lorentz symmetry. 

For two and three extra dimensions, we saw that the way that the resulting vectors of $O(s_2)$ couple with the gauge fields allows us to identify the gauge fields as those of $U(1)$ and $SU(2)$ symmetries respectively. The remaining components of a higher-dimensional vector form a Lorentz four-vector, which couples to gravity (and inertial forces) in the way expected from GR. (That is, through the usual four-dimensional Levi-Civita connection.)

We have seen that when these gauge fields have zero field strength, the product space becomes a Cartesian product space. On such a space, the gauge fields can only ever be artefacts of choosing a coordinate system which does not respect the factor spaces.

The classical vacuum of such a theory has this feature that any gauge fields have zero field strength. However, it also has zero Riemann curvature on the four-dimensional factor spacetime, and the compact space is a product of Einstein manifolds and two-dimensional manifolds. This represents deep space, in the absence of passing gravitational waves or gauge field waves.

We have also investigated the links between the geometry and group theory of product spaces and the orbits formed by index-aligned tensors under changes of coordinate basis. In particular, we have shown that whenever \emph{any} index-aligned tensor can be diagonalised across a neighbourhood, such that it has the same multiplicities of eigenvalues at every point on the neighbourhood, the spacetime is isometric to a product manifold across this neighbourhood. The multiplicities of its eigenvalues are the dimensions of the factor spaces and these are determined by its algebraic invariants. This enables us to find particular forms for the algebraic invariants which correspond to particular Kaluza-Klein theories.

Finally, we have a specific case of this result relating to the classical vacuum of a Kaluza-Klein theory. This occurs when the \emph{Ricci} tensor can be diagonalised across a neighbourhood, such that its eigenvalues corresponding to the four-dimensional spacetime are all zero and the remaining eigenvalues occur in multiplicities appropriate to the gauge fields. These eigenvalues are determined by the pure Ricci invariants and are constrained by the contracted Bianchi identity to be constant, apart from those which relate to a two-dimensional factor space.

\medskip

Now, it may reasonably pointed out that the unification prescription given here embeds four-dimensional spacetime symmetries and other symmetries (related to internal symmetries) in a larger group. This appears to violate several no-go theorems of the 1960s, the most famous and comprehensive of which was that of Coleman and Mandula\cite{CM1}. It is at present difficult to evaluate whether the Coleman-Mandula theorem is a serious problem for the kinds of Kaluza-Klein theories envisaged here. This is because the theorem deals with the symmetries of multi-particle interactions, when single-particle field excitations have yet to be studied in the models considered here, let alone interactions between them.

However, its immediate predecessor, O'Raifeartaigh's no-go theorem\cite{O'R1}, is based purely on issues relating to Lie algebras and their corresponding operators on field states. In a forthcoming paper, we will explain in detail how the kinds of Kaluza-Klein theories put forward here take advantage of a loophole in the theorem. However, in outline, the argument is this. 

In his paper, O'Raifeartaigh found four ways in which the Lie algebra of the Poincar\'{e} group and the Lie algebra of an internal symmetry group could be embedded in a larger Lie algebra. The only one of the four ways that he considered did not have unphysical aspects was if that larger Lie algebra was simply the direct sum of those of the Poincar\'{e} and internal symmetry groups.

Now our embedding of $so(1,3)$ and $so(s_2)$ into $so(1, 3 + s_2)$ corresponds to his `case (ii)'. His issue with this was that it implied a higher-dimensional algebra of translations - the operators for which would be mutually commuting and have continuous spectra of eigenstates.

In our case, this higher-dimensional translation algebra becomes apparent in the decompactification limit. In this limit, we obtain a flat $N$-dimensional spacetime. We can then consider an $N$-dimensional translation operating on any scalar, vector or tensor field in this spacetime, displacing a chosen field configuration by amounts $\delta x^I$ in each direction. We can identify eigenstates under this operation. We find that the operators corresponding to the translations through $\delta x^Y$ are indeed mutually commuting, commute with the four-dimensional translation operators and have continuous spectra of eigenstates.

However, when we start to compactify, we replace our $\mathbb{R}^{N_2}$ subspace with a compact space. If, for example, this is a spherical space $S^{N_2}$, we are taking its radius from infinity down to a finite value. Then the extra translations become replaced by transformations on the compact space - on $S^{N_2}$ these are rotations that do not belong to the $H_2$ subgroup. Such transformations do not commute - with each other, or indeed with the operators of the gauge group $H_2$. They also have discrete eigenstates, evading the problem spotted by O'Raifeartaigh.

\medskip

It is worth noting that the main results of this paper do not depend on any specific form of field equations or action integral. This shows us how much can be deduced purely from considering symmetry transformation groups. However, field equations are needed to describe how the field content varies from one event in spacetime to another. This will also explored in a forthcoming paper.

\medskip

The third area where there is a gap in the above analysis is the inclusion of spinors. We have indicated above that they can be combined to construct vector fields, but we have not described the spaces they inhabit, nor how those spaces relate to those described in this paper. Further research is needed on this, but clearly Clifford algebra structures will play a key role. 

We can see this by thinking about odd numbers of additional dimensions. These lead to gauge groups of the form $SO(2n + 1)$, which has a spinor representation with $d$ complex components, where $d = 2^n$. Thus when $n=1$, the gauge group is $SO(3)$, which has a two-component spinor; when $n=2$, the gauge group is $SO(5)$, which has a four-component spinor; when $n=3$, the gauge group is $SO(7)$, which has a eight-component spinor and so forth. The outer product of a spinor and its adjoint spans a $d^2$-dimensional space and forms a representation of $U(d)$. This space is also spanned by the matrices of the Clifford algebra structure of the gauge group, which are created by taking alternating commutators and anticommutators of its $\gamma$-matrices. We can use the $N = 2n+1$ $\gamma$-matrices themselves to project the components of a vector field out of the outer product.

This needs to be understood, for example, to handle $SU(3)$ as an internal symmetry group, as it would need to be embedded in $U(4)$ or a larger $U(d)$ group.

\section*{Acknowledgements}

This paper is dedicated to the memory or Prof. K. J. Barnes. A wonderfully jolly and inspirational PhD supervisor, he introduced me to the fascinating world of non-linear realisations and set me looking at a symmetry breaking pattern which underlies the first example in this paper. I would particularly like to thank Iuval Clejan for many probing questions on this paper and for helping me identify some minor inaccuracies and useful clarifications. I would also like to thank David McNutt for a useful and thought-provoking Zoom call. My thanks also go to Janilo Santos, Malcolm MacCallum and Joana Cirici as well as my colleagues at the Centre for Advancing Mathematics and Physics at the Ronin Institute for making helpful comments on the draft paper and for directing me to relevant papers. Finally, I would like to thank the creators of \href{https://matrixcalc.org/}{Matrix calculator} and 
\href{https://matrix.reshish.com/}{Matrix Reshish}, whose tools have helped me gain insights into tensor orbits, as well as all those who put helpful postings on sites such as StackExchange and ResearchGate and those who have edited Wikipedia pages on these topics.

\appendix

\section{Appendix - More on diagonalisation and Cartan subspaces \label{Appx}}

\subsection{Diagonalisation when the metric is or is not positive definite}

We saw in Section \ref{Dual} that for any tangent space $T_A \mathcal{M}$, we can always define a set of Riemann normal coordinates whose basis is orthonormal on this tangent space. This allows us to cast any rank-two tensor $X_I{}^J$ in operator form into a frame basis.

For a space of positive definite signature, in this frame basis, indices are raised and lowered using $\delta^{IJ}$ and $\delta_{IJ}$. Let us call a chosen set of Riemann normal coordinates $x'$. We then find that for any index-aligned tensor,
\begin{equation}
	X^{(x')}_I{}^J = X_{(x')}^J{}_I = \delta^{JK} X^{(x')}_K{}^L \delta_{LI} 
\end{equation}
so, for example,
\begin{equation}
	X^{(x')}_1{}^2 = \delta^{2K} X^{(x')}_K{}^L \delta_{L1} = X^{(x')}_2{}^1
\end{equation}
We therefore see that $X^{(x')}$ is always symmetric. The spectral theorem tells us that any symmetric matrix may be diagonalised using an orthogonal transformation. Thus on a space with positive definite metric, if $X^{(u)}_I{}^J$ is any index-aligned tensor in \emph{any} coordinate system, there is always a change of coordinates which will diagonalise it\cite{HC}.

We have shown that any such change of coordinates $j_0$ can always be decomposed in the form (\ref{jLg}). This means that $X$ can be diagonalised by $L_0$, a representative of the coset space $J/\mathrm{Stab}_X$. 

Indeed, we can use a further decomposition to describe the above two-stage diagonalisation. As we explained in Section \ref{Sec-jli}, any element of $j_0$ which maps a coordinate basis to a frame basis can be decomposed in the form (\ref{jli}), where $i_0$ is an orthogonal or pseudo-orthogonal transformation. The $L_0$ which diagonalises $X$ can be decomposed in the same way:
\begin{equation}
	L_0 = \tilde{l}_0 \, \tilde{i}_0
\end{equation}
where $\tilde{i}_0$ is an orthogonal transformation. $\tilde{l}_0$ then represents a transformation which puts $X^{(u)}$ into a frame basis $\hat{\mathbf{k}}_I$, associated with the Riemann normal coordinates $x'$. $\tilde{i}_0$ carries out the second stage of the diagonalisation, transforming it into a second frame basis, $\hat{\mathbf{n}^I}$.

\medskip

If the metric is indefinite, we use $\eta^{IJ}$ and $\eta_{IJ}$ to raise and lower indices in the $\hat{\mathbf{k}}$-frame instead of $\delta^{IJ}$ and $\delta_{IJ}$. In this case, $X^{(x')}$ is not symmetric. Instead, 
\begin{equation}
	X^{(x')}_I{}^J = X_{(x')}^J{}_I = \eta^{JK} X^{(x')}_K{}^L \eta_{LI} 
\end{equation}
so that in a Lorentzian spacetime, for example,
\begin{equation}
	X^{(x')}_0{}^2 = X_{(x')}^J{}_I = \eta^{2K} X^{(x')}_K{}^L \eta_{L0} 
\end{equation}
We then find that in general, $X^{(x')}$ takes the form
\begin{equation} \label{ABBC}
	X^{(x')} = \left( \begin{array}{cc}
		A & B \\
		-B^{\mathrm{T}} & C
	\end{array} \right)
\end{equation}
where $A$ and $C$ are symmetric matrices. $i_0$ is then a pseudo-orthonormal transformation, as described in \cite{TSS}. This cannot always be used to diagonalise $X$, as described in Section \ref{diag-non}.

\subsection{Proof that any index-aligned tensor with distinct real eigenvalues may be diagonalised using an element of $GL(N, \mathbb{R})$}

While not all rank-two tensors can be diagonalised using $J$ on a spacetime with indefinite metric, some can. We now show that, regardless of the signature of the metric, any index-aligned tensor with distinct real eigenvalues may be diagonalised using an element of $GL(N, \mathbb{R})$.

\medskip

If a real eigenvalue $\lambda$ of any real index-aligned tensor $X$ has a complex eigenvector $t^I = v^I + \mathrm{i} w^I$, the real part $v^I$ is also an eigenvector, because if 
\begin{equation}
	X^I_J (v^I + \mathrm{i} w^I) = \lambda (v^I + \mathrm{i} w^I), 
\end{equation}
then
\begin{equation}
	X^I_J v^I = \lambda v^I
\end{equation}
by equating real components. 

For any real index-aligned tensor with \emph{distinct} real eigenvalues, each real eigenvector is then specified up to a real scale factor. We will consider normalisation of these eigenvectors below. For now, let us denote a set of real eigenvectors of an index-aligned tensor $X$ with distinct eigenvalues as $v^I{}_{(0)}, v^I{}_{(1)}, \ldots, v^I{}_{(N)}$, and their respective eigenvalues $\lambda_{(0)}, \lambda_{(1)}, \ldots, \lambda_{(N)}$. Then, for example, the eigenvalue equation for $v^I{}_{(1)}$ reads in full:
\begin{equation}
	\left( \begin{array}{cccc}
		X^0_0 & X^0_1 & \ldots & X^0_{N} \\
		X^1_0 & X^1_1 & & \\
		\vdots &  & \ddots & \\
		X^{N}_0 & & & X^{N}_{N}
	\end{array} \right)
	\left( \begin{array}{c}
		v^0_{(1)} \\
		v^1_{(1)} \\
		\vdots \\
		v^{N}_{(1)}
	\end{array} \right)
	= \lambda_{(1)}
	\left( \begin{array}{c}
		v^0_{(1)} \\
		v^1_{(1)} \\
		\vdots \\
		v^{N}_{(1)}
	\end{array} \right)
\end{equation}
Note that:
\begin{itemize}
	\item again, we use the convention that the indices run $0,1,2,3,5,6, \ldots$, $N$
	\item \emph{in the above and what follows, whenever an index in brackets is repeated, it is assumed \emph{not} to be summed, unless stated explicitly with a summation symbol}. 
\end{itemize}

It is then easy to show that these eigenvectors are mutually orthogonal and linearly independent. For any two real eigenvectors,
\begin{equation}
	v_{I (K)} X^I_J v^J{}_{(L)} = \lambda_{(L)} v_{I (K)} v^I{}_{(L)}
\end{equation}
but also
\begin{equation}
	v_{I (K)} X^I_J v^J{}_{(L)} = v^J{}_{(L)} X^I_J v_{I (K)} 
	= v_{J (L)} X_I^J v^I{}_{(K)} = \lambda_{(K)} v_{I (K)} v^I{}_{(L)}
\end{equation}
Thus
\begin{equation}
	(\lambda_{(K)} - \lambda_{(L)}) \, v_{I (K)} v^I{}_{(L)} = 0
\end{equation}
Then if $K \neq L$, as the eigenvalues are distinct,
\begin{equation}
	v_{I (K)} v^I{}_{(L)} = 0
\end{equation}
This means that all the eigenvectors are mutually orthogonal. There are $N$ of them in a $N$-dimensional space, so they must be linearly independent, as claimed.

From this result, it is easy to show that any such tensor can be diagonalised using a matrix $E^I{}_J \in J_A$ composed of its real eigenvectors. This matrix $E$ is defined as having elements $E^I{}_J = v^I{}_{(J)}$:

\begin{equation}
	E = \left( \begin{array}{cccc}
		v^0{}_{(0)} & v^0{}_{(1)} & \ldots & v^0{}_{(N)} \\
		v^1{}_{(0)} & v^1{}_{(1)} & & \\
		\vdots &  & \ddots & \\
		v^{N}{}_{(0)} & & & v^{N}{}_{(N)}
	\end{array} \right)
\end{equation}
so that each column is an eigenvector. Then
\begin{equation} \label{evec-mtx}
	X^I_J E^J{}_K = X^I_J v^J{}_{(K)} = \lambda_{(K)} v^I{}_{(K)} = \lambda_{(K)} E^I{}_K .
\end{equation}

Now the eigenvectors are linearly independent, so $|E| \neq 0$, so $E$ is invertible. We can therefore multiply (\ref{evec-mtx}) by $E^{-1}$:
\begin{equation}
	(E^{-1})^L{}_I X^I_J E^J{}_K = \lambda_{(K)} \delta^L_K
\end{equation}
thus
\begin{equation} \label{X-diag}
	E^{-1} X E = \begin{pmatrix}
		\lambda_{(0)} & &  & \\
		& \lambda_{(1)} & & \\
		&  & \ddots & \\
		& & & \lambda_{(N)}
	\end{pmatrix}
\end{equation}
Thus $E$ diagonalises $X$. It is composed of real eigenvectors, so all of its elements are real. It is also invertible. It is therefore an element of $J_A$ as claimed. This means that (\ref{X-diag}) is an inner automorphic mapping.

\subsection{Interpretation of this diagonalisation}

This can be interpreted as follows\cite{Boas}. The action of $X$ on the value of a vector field at $A$ in $u$-coordinates is
\begin{equation} \label{X-on-V}
	X^I{}_K: \mathbf{V} |_A = V_{(u)}^I |_A \, \mathbf{e}_I |_A 
	\mapsto \mathbf{V'} |_A = V'^I_{(u)} |_A \, \mathbf{e}_I |_A = X^I{}_K V^K_{(u)} |_A \, \mathbf{e}_I |_A
\end{equation}
Then $v^I{}_{(0)} \mathbf{e}_I |_A, v^I{}_{(1)} \mathbf{e}_I |_A, \ldots$ are a set of $N$ vectors which are eigenvectors of the transformation, that is, they are scaled by $\lambda_0, \lambda_1, \ldots$ respectively under (\ref{X-on-V}). These can be viewed as an alternative basis on $T_A \mathcal{M}$:
\begin{equation} \label{evec-basis}
	\mathbf{e}'_0 |_A = v^I{}_{(0)} \mathbf{e}_I |_A, \; \mathbf{e}'_1 |_A = v^I{}_{(1)} \mathbf{e}_I |_A, \; \ldots
\end{equation}
This change of basis may then be written
\begin{equation}
	\mathbf{e}'_J |_A = E^I{}_J \, \mathbf{e}_I |_A 
	\; \Rightarrow \; \mathbf{e}_I |_A = (E^{-1})^K{}_I \, \mathbf{e}'_K |_A
\end{equation}
with the corresponding transformation of the vector components being:
\begin{equation}
	V_{(u)}^K |_A = V^L_{(u')} |_A \, E^K{}_L
\end{equation}
By substituting these into (\ref{X-on-V}), we can find the action of $X$ on $V$ in the $u'$-coordinate system:
\begin{equation}
	X: V^K_{(u')} |_A \mapsto V'^K_{(u')} |_A = (E^{-1} X E)^K{}_L V^L_{(u')} |_A
\end{equation}
Thus the matrix
\begin{equation}
	D \equiv E^{-1} X E
\end{equation}
is $X$ in $u'$ coordinates, which acts on $V^K_{(u')} |_A$. It simply scales the components:
\begin{equation}
	V'^K_{(u')} |_A = 
	\begin{pmatrix}
		\lambda_{(0)} & &  & \\
		& \lambda_{(1)} & & \\
		&  & \ddots & \\
		& & & \lambda_{(N)}
	\end{pmatrix}
	\begin{pmatrix}
		V^0_{(u')} \\
		V^1_{(u')} \\
		\vdots \\
		V^N_{(u')}
	\end{pmatrix}
	= \begin{pmatrix}
		\lambda_{(0)} V^0_{(u')} \\
		\lambda_{(1)} V^1_{(u')} \\
		\vdots \\
		\lambda_{(N)} V^{N}_{(u')}
	\end{pmatrix}
\end{equation}

We can use (\ref{evec-basis}) to find the metric at $A$ for the new basis:
\begin{equation} \label{diag-metric}
	\mathrm{g}^{(u')}_{KL} |_A = (\mathbf{e}'_K, \mathbf{e}'_L)_A = v^I{}_{(K)} v^J{}_{(L)} (\mathbf{e}_I, \mathbf{e}_J)_A = v^I{}_{(K)} v^J{}_{(L)} \mathrm{g}_{IJ} |_A  = v^I{}_{(K)} v_{I (L)}
\end{equation}
By choosing $K \neq L$, we see that the new basis is an orthogonal one. Each basis vector can then be scaled to give us a pseudo-orthonormal basis, with $t$ timelike basis vectors and $s$ spacelike ones.

Indeed, we can obtain an orthonormal basis directly from (\ref{evec-basis}) if we choose the eigenvectors appropriately. If we take any non-null set of vector components $v^I$, we may define a normalised vector by 
\begin{equation}
	\hat{v}^I = \dfrac{v^I}{\sqrt{|v^J v_J|}}
\end{equation}
Then
\begin{equation}
	\hat{v}^I \hat{v}_I = \dfrac{v^I v_I}{|v^J v_J|}
\end{equation}
which is $+1$ or $-1$ according to whether $v$ is timelike or spacelike. If $v$ is an eigenvector of $X$, $\hat{v}$ is also an eigenvector, with the same eigenvalue as $v$:
\begin{equation}
	X_I^K \hat{v}^I = \dfrac{1}{\sqrt{|v^J v_J|}} X_I^K v^I  = \dfrac{1}{\sqrt{|v^J v_J|}} \lambda v^K = \lambda \hat{v}^K
\end{equation}
Now normalise each of the eigenvectors of $X$ in this way. Let $\hat{E}$ be the matrix formed from these eigenvectors, so that
\begin{equation}
	\hat{E}^I{}_J = \hat{v}^I{}_{(J)}
\end{equation}
Then (\ref{diag-metric}) becomes
\begin{equation}
	\mathrm{g}^{(u')}_{KL} |_A = \hat{v}^I{}_{(K)} \hat{v}_{I (L)} = \eta_{KL}
\end{equation}
so that the basis defined by 
\begin{equation}
	\hat{\mathbf{n}}_0 |_A = \hat{v}^I{}_{(0)} \mathbf{e}_I |_A, \; \hat{\mathbf{n}}_1 |_A = \hat{v}^I{}_{(1)} \mathbf{e}_I |_A, \; \ldots
\end{equation}
is pseudo-orthonormal.

\subsection{More on Cartan subspaces}

If $Y$ is an element of $gl(N,\mathbb{R})$ which commutes with another element $X$ which has distinct eigenvalues, $Y$ lies in the Cartan subalgebra containing $X$. This means that it can be expressed as a linear sum of the powers of $X$:
\begin{equation}
	Y = \alpha \mathbf{1} + \beta X + \gamma X^2 + \ldots
\end{equation}
It is easy to see that this implies that $Y$ is diagonalised using the same matrix $E$ of the eigenvectors of $X$ that diagonalises $X$, as follows. Acting with $Y$ on any eigenvector $v_{(J)}$ of $X$ gives us
\begin{equation}
	Y v_{(J)} = \alpha \mathbf{1} v_{(J)} + \beta X v_{(J)} + \gamma X^2 v_{(J)} + \ldots
\end{equation}
Then using the eigenvalue equation (\ref{eval}) for $v_{(J)}$, we get
\begin{equation}
	Y v_{(J)} = \alpha v_{(J)} + \beta \lambda v_{(J)} + \gamma \lambda^2  v_{(J)} + \ldots = (\alpha + \beta \lambda + \gamma \lambda^2 + \ldots) \, v_{(J)} 
\end{equation}
Thus $v_{(J)}$ is an eigenvector of $Y$, with eigenvalue $\alpha + \beta \lambda + \gamma \lambda^2 + \ldots$ . The set of eigenvectors $\{ \hat{v}_{(J)} \}$ then form a set of mutually orthonormal, linearly independent eigenvectors for $Y$, so the matrix $E$ constructed from them diagonalises $Y$.

If $Y$ has distinct eigenvalues, it is only diagonalised by matrices built from eigenvectors of tensors in that single Cartan subspace. If it has repeated eigenvalues, then it is located in an infinite number of Cartan subspaces, and we can express it in terms of matrices spanning any of these Cartan subspaces. Any such expansion can be used to diagonalise it.

Cartan subspaces may also be spanned by a set of projection operators. These can be used for calculating finite elements of $J_A$ from the tensors that generate them (if they are diagonalisable), in the same way that Rosen\cite{Rosen} and Barnes \emph{et al}\cite{BDS1} did for $SU(3)$. For example, the diagonal Cartan subspace is spanned by
\begin{eqnarray}
	P_{(0)} = \begin{pmatrix}
		1 & &  & & \\
		& 0 & & & \\
		&  & 0 & & \\
		&  & & \ddots & \\
		& & & & 0
	\end{pmatrix}, 
	& & P_{(1)} = \begin{pmatrix}
		0 & &  & & \\
		& 1 & & & \\
		&  & 0 & & \\
		&  & & \ddots & \\
		& & & & 0
	\end{pmatrix}, \nonumber \\
	& & P_{(2)} = \begin{pmatrix}
		0 & &  & & \\
		& 0 & & & \\
		&  & 1 & & \\
		&  & & \ddots & \\
		& & & & 0
	\end{pmatrix},  \; \; \ldots 
\end{eqnarray}
These have the properties
\begin{equation} \label{po-cond1}
	P_{(K)} P_{(L)} = \left\{ \begin{array} {ll} 
		0 & \mbox{if $K \neq L$} \\
		P_{(K)} & \mbox{if $K = L$}
	\end{array}
	\right.
\end{equation}
and
\begin{equation} \label{po-cond2}
	\sum \limits_K P_{(K)} = \mathbf{1}
\end{equation} 
A diagonal matrix may then be written
\begin{equation}
	D = \sum_K \lambda_{(K)} P_{(K)}
\end{equation}

Under the inner automorphism,
\begin{equation}
	E: D^I_J \rightarrow X^I_J = \sum_K \lambda_{(K)} (E P_{(K)} E^{-1})^I_J
\end{equation}
Note that the set of matrices
\begin{equation}
	P'_{(K)} = E P_{(K)} E^{-1}
\end{equation}
also satisfy (\ref{po-cond1}) and (\ref{po-cond2}) - that is, they are also a complete set of projection operators, this time for the Cartan subspace containing $X$. (When $X$ is a curvature operator which is diagonalisable, these appear to be the curvature projectors used by Hervik and Coley\cite{HC}.) This means that if $X$ generates an element $j$:
\begin{equation}
	j = \mathrm{e}^{\mathrm{i} X}
\end{equation}
then
\begin{equation}
	j =  \sum_K P'_{(K)} \mathrm{e}^{\mathrm{i} \lambda_{(K)}}
\end{equation}

Note also that if $E$ is pseudo-orthogonal, 
\begin{equation}
	(E^{-1})^I{}_J = \eta^{IK} (E^{-1})_K{}^L \eta_{LJ} = \eta^{IK} E^L{}_K \eta_{LJ} = \eta^{IK} v^L{}_{(K)} \eta_{LJ}
\end{equation}
so that
\begin{equation}
	(P'_{(K)})^I{}_P = v^I{}_{(J)} (P_{(K)})^J{}_L \eta^{LM} v^N{}_{(M)} \eta_{NP}
\end{equation}
so for $K$ a timelike direction, we have
\begin{equation}
	(P'_{(K)})^I{}_P = v^I{}_{(K)} v^N{}_{(K)} \eta_{NP}
\end{equation}
while for $K$ a spacelike direction,
\begin{equation}
	(P'_{(K)})^I{}_P = - v^I{}_{(K)} v^N{}_{(K)} \eta_{NP}
\end{equation}
- the projection operators for a diagonalisable tensor are constructed from its own eigenvectors.

\end{document}